\pdfoutput=1
\documentclass[fleqn,usenatbib]{mnras}

\usepackage{newtxtext,newtxmath}

\usepackage[T1]{fontenc}
\usepackage{ae,aecompl}

\usepackage{graphicx}	
\usepackage{amsmath}	
\usepackage{amssymb}	
\usepackage{float}
\usepackage{gensymb}
\usepackage{tabu}
\usepackage{enumerate}
\title[Size, shade or shape?]{Size, shade or shape? The contribution of galaxies of different types to the star-formation history of the Universe from \textit{SDSS-IV~MaNGA}}

\author[T. Peterken et al.]{
Thomas Peterken,$^{1}$\thanks{E-mail: Thomas.Peterken@nottingham.ac.uk}
Alfonso Arag{\'o}n-Salamanca,$^{1}$\thanks{E-mail: Alfonso.Aragon@nottingham.ac.uk}
Michael Merrifield,$^{1}$
\newauthor Vladimir Avila-Reese,$^{2}$ Nicholas F.\ Boardman,$^{3}$ Helena Dom{\'i}nguez S{\'a}nchez,$^{4}$
\newauthor Dmitry Bizyaev,$^{5}$ Niv Drory,$^{6}$ Kaike Pan,$^{5}$ Joel R.\ Brownstein$^{7}$\\
$^{1}$School of Physics and Astronomy, University of Nottingham, University Park, Nottingham NG7 2RD, UK\\
$^{2}$Instituto de Astronom{\'i}a, Universidad Nacional Aut{\'o}noma de M{\'e}xico, A.P. 70--264, 04510 CDMX, M{\'e}xico\\
$^{3}$Department of Physics and Astronomy, University of Utah, Salt Lake City, UT 84112, USA\\
$^{4}$Institute of Space Sciences (ICE, CSIC), Campus UAB, Carrer de Magrans, E-08193 Barcelona, Spain\\
$^{5}$New Mexico State University, Apache Point Observatory, P.O. Box 59, Sunspot, NM 88349\\
$^{6}$McDonald Observatory, The University of Texas at Austin, 1 University Station, Austin, TX 78712, USA\\
$^{7}$Department of Physics and Astronomy, University of Utah, 115 S. 1400 E., Salt Lake City, UT 84112, USA
}

\date{Draft copy: \today}

\pubyear{2020}

\begin{document}
\label{firstpage}
\pagerange{\pageref{firstpage}--\pageref{lastpage}}
\maketitle

\begin{abstract}
By fitting stellar populations to SDSS-IV MaNGA survey observations of $\sim7000$ suitably-weighted individual galaxies, we reconstruct the star-formation history of the Universe, which we find to be in reasonable  agreement with previous studies.  Dividing the galaxies by their present-day stellar mass, we demonstrate the downsizing phenomenon, whereby the more massive galaxies hosted the most star-formation at earlier times.  Further dividing the galaxy sample by colour and morphology, we find that a galaxy's present-day colour tell us more about its historical contribution to the cosmic star formation history than its current morphology.  We show that downsizing effects are greatest among galaxies currently in the blue cloud, but that the level of downsizing in galaxies of different morphologies depends quite sensitively on the morphological classification used, due largely to the difficulty in classifying the smaller low-mass galaxies from their ground-based images.  Nevertheless, we find agreement that among galaxies with stellar masses $M_{\star}>6\times10^{9}\,M_{\odot}$, downsizing is most significant in spirals.  However, there are complicating factors.  For example, for more massive galaxies, we find that colour and morphology are predictors of the past star formation over a longer timescale than in less massive systems.  Presumably this effect is reflecting the longer period of evolution required to alter these larger galaxies' physical properties, but shows that conclusions based on any single property don't tell the full story.
\end{abstract}




\section{Introduction}
\label{sec:Intro}

The question of when and where the stars residing in today's galaxies formed is essential to understanding the present-day Universe.  Since early works by \citet{Madau+96}, \citet{Connolly+97} and others, many studies have measured the average instantaneous star-formation rate in galaxy populations observed at different redshift snapshots to build a picture of the overall cosmic star-formation history; see \citealt{MD14} for a comprehensive review.  These approaches have built up a picture of a two-phase star-formation history of the Universe, with the mean star-formation rate per unit of comoving volume rising rapidly during the first $1-2\,\textrm{Gyr}$ (i.e.\ $z>4$) after the Big Bang, and declining in more recent times.  The exact lookback time at which the peak of cosmic star-formation occurred is uncertain, but most sources agree that it is broadly in the range of $z=1-3$, corresponding to approximately $8-11\,\textrm{Gyr}$ before the present day; see analyses by \citet{HopkinsBeacom06}, \citet{Behroozi+13}, and \citet{MD14}, and references therein.

Within this picture, studies of galaxy populations at different redshifts have demonstrated the link between the star-formation history of the Universe and the evolution of its constituent galaxies.  For example, many show that the star-formation rate of galaxies with high stellar mass peaked and declined at preferentially earlier times than low-mass galaxies \citep{Cowie+96, Fontanot+09, Peng+10, Muzzin+13}, an effect normally referred to as ``downsizing''.  Others have proven that a link between a galaxy's morphology and its contribution to the cosmic star-formation history exists over a range of redshifts \citep{Wuyts+11, Bell+12, Cheung+12, Mortlock+13, Moresco+13}, although determining detailed morphologies at high redshift is of course difficult.

It has long been known that in the present-day Universe, spirals are generally bluer in colour (e.g.\ \citealt{Holmberg58}) due to their higher rate of star formation (e.g.\ \citealt{Roberts63, Kennicutt83}), and are less massive \citep{BlantonMoustakas09} than earlier-type galaxies, suggesting an evolutionary sequence of galaxies transitioning from late to early types as they grow and subsequently cease their star-formation.  However, the exact mechanisms by which galaxies are able to alter their shape and colour are still not fully understood.  One avenue to studying the typical evolution processes which have occurred in galaxies could be to determine whether a present-day galaxy's colour\footnote{``Colour'' here refers to a broadband measure of the spectral energy distribution in the optical region} or its morphology is the strongest indicator to its past.  It would then be possible to understand which property transition --- morphology or colour; ``shape'' or ``shade'' --- is more fundamental, and whether the associated timescales vary according to galaxy properties such as stellar mass.

Unfortunately, by their nature, studies of galaxy populations at different redshifts are limited to studying the average statistical behaviour of a galaxy population's star-formation at different snapshots in the Universe's history, and are therefore unable to trace how individual galaxies have evolved.  In order to understand the link between a galaxy's \textit{present-day} properties and its past contribution to the cosmic star-formation history, an alternative approach is therefore required.

Using large samples of galaxies from spectroscopic surveys, \citet{PanterHeavensJimenez03, Panter+07} and \citet{Heavens+04} showed that by measuring galaxies' individual star-formation histories using spectral fitting techniques, the star-formation history of the Universe can be reconstructed.  Despite significant progress in more recent years, obtaining accurate non-parametric star-formation histories of galaxies in this manner is still subject to significant uncertainties, particularly those due to difficulties in creating reliable stellar population templates using stellar evolution models which are still poorly understood \citep{Charlot+96, Maraston98, Maraston05, Yi03, Lee+07, Ge+19}, and also due to assumptions about the behaviour of the stellar initial mass function \citep{Maraston98, vanDokkum+08, Pforr+12, CidFernandes+14} and the treatment of stellar elemental abundances.  See \citet{Conroy+09, Conroy+10}, and \citet{ConroyGunn10} for a set of detailed reviews and discussion on this important subject.

However, notwithstanding these difficulties, many studies have shown that non-parametric stellar population fitting methods produce generally reliable results (see e.g.\ \citealt{Starlight, Panter+07, Pipe3D, Li+17, deAmorim+17, Ge+18, FakeNews}; see also Appendix~A of \citealt{Peterken+20FR} for tests specific for the spectral fitting methods used here), and these stellar~population ``fossil~record'' methods have therefore been successfully applied to modern integral-field spectroscopic galaxy surveys to uncover the history of low-redshift galaxies and their physical components (see e.g.\ \citealt{CidFernandes+13, Perez+13, IbarraMedel+16, GonzalezDelgado+16, GonzalezDelgado+17, Peterken+19TS, Peterken+20FR, GarciaBenito+19, FraserMcKelvie+19}).  Others have used spectral fitting analyses to show that a galaxy's observed colour (e.g.\ \citealt{IbarraMedel+16}; or equivalently star-formation rate, e.g.\ \citealt{Sanchez+19}) and morphology (e.g.\ \citealt{GarciaBenito+17, LopezFernandez+18, Lacerna+20, Bellstedt+20}) are directly linked to the historical evolution of its star formation rate.  In a cosmological context, \citet{LopezFernandez+18} and \citet{Sanchez+19} have demonstrated the power of using fossil record techniques to reconstruct how galaxies' physical properties have evolved over cosmic time, showing good agreement with redshift snapshot studies, and thereby justifying this approach as a complementary analysis technique to study the connection between the Universe's evolution with its present-day galaxies.

With its consistent radial coverage of a large sample of galaxies with varying physical properties, the integral-field spectroscopic MaNGA survey \citep{Bundy+15} (part of the fourth generation of the Sloan Digital Sky Survey; SDSS-IV; \citealt{Blanton+17}) offers an ideal tool to investigate the link between today's galaxies and the Universe's past.  In \citet{Peterken+20Morph}, we explored how the stellar population fossil~record can reveal the cosmic evolution of the star-formation ``main~sequence'' and the mass function of galaxies.  Here, we use the same measured star-formation histories of a large sample of galaxies to derive the star-formation history of the Universe, and make use of morphological information from both citizen science and machine learning classifications to explore the connection between present-day stellar mass, colour, and morphology to a galaxy's star-formation rate evolution over the age of the Universe.

\bigskip

This paper is structured as follows.  In Section~\ref{sec:Data-MaNGA}, we outline the relevant details of the SDSS-IV MaNGA survey.  We outline how the samples were selected and describe the division of these samples into morphological and colour sub-samples in Section~\ref{sec:Sample}.  We then briefly summarise the spectral fitting methods we employ to measure galaxy star-formation histories in Section~\ref{sec:Fitting}.  Section~\ref{sec:SFHoU} contains the derivation of the cosmic star-formation history and its contribution from galaxies of different present-day stellar masses (\ref{sec:SFHoU-Mass}), colour (\ref{sec:SFHoU-Col}), and morphological classifications (\ref{sec:SFHoU-Morph}).  Finally, we interpret how these results fit into a context of downsizing and discuss the relative importance of morphology and broadband colour in Section~\ref{sec:Interpretation}.

\medskip

Throughout this paper, we assume a flat~$\Lambda$CDM cosmology with $H_0=68\,\textrm{km}\,\textrm{s}^{-1}\,\textrm{Mpc}^{-1}$ and $\Omega_{\rm m}=0.308$, consistent with \citet{Planck15}.

\section{MaNGA}
\label{sec:Data-MaNGA}

As part of the SDSS-IV, Mapping Nearby Galaxies at Apache Point Observatory (MaNGA; \citealt{Bundy+15}) is an integral-field spectroscopic galaxy survey.  As of the time of writing, on-site operations have been completed, and by the end of 2020 the fully-reduced observations will be available for over 10,000 low-redshift ($0.01<z<0.15$, median $z\sim0.3$) galaxies with a spatial resolution of 2.5~arcseconds \citep{Yan+16-design}.  Observations make use of specially-designed integral field units of five sizes ranging from 12 to 32~arcsecond diameters with 19 to 127~fibres \citep{Drory+15}, which are mounted onto plates on the 2.5-metre Sloan telescope at Apache Point Observatory in New Mexico \citep{Gunn+06}.  The fibres are fed into the BOSS spectrographs \citep{Smee+13} and the spectra are calibrated to better than 5\% accuracy \citep{Yan+16-cal} covering a wavelength range of $3600-10300\,\textrm{\AA}$ with a resolution of $R\approx2000$.  Observations are designed to reach a minimum signal-to-noise ratio of $5\,\textrm{\AA}^{-1}$ at $1.5\,R_{\rm e}$ \citep{Law+15}, where $R_{\rm e}$ is the effective radius of each observed galaxy measured by the NASA-Sloan Atlas (NSA; \citealt{NSA}).

The calibrated spectra are reduced and combined into three-dimensional datacubes by a custom data reduction pipeline (DRP; \citealt{Law+16}; Law et al.\ in preparation), and a data analysis pipeline (DAP; \citealt{DAP, Marvin}) provides data analysis products such as spectral index maps, stellar and gas kinematics, and emission line fluxes \citep{DAPLines}.

\section{Sample selection}
\label{sec:Sample}

MaNGA targets galaxies with a flat distribution in log(stellar~mass) over the range of $10^{9} M_{\odot} < M_{\star} < 10^{11.5} M_{\odot}$, and the full targeting catalogue assigns galaxies to designated samples.  The Primary sample galaxies are observed to a radius of $1.5\,R_{\rm e}$, while the Secondary sample contains observations to $2.5\,R_{\rm e}$.  The Primary sample is supplemented by a colour-enhanced sample to form the Primary+ sample, which over-samples unusual regions of the stellar mass--colour plane such as high-mass blue galaxies, low-mass red galaxies, and the ``green~valley'' \citep{Law+15}.

From the latest internal MaNGA data release (MaNGA Product Launch 9; MPL-9), we selected all galaxies belonging to any of the Primary, Primary+, or Secondary samples.  In doing so, we required the MaNGA DRP to have assigned no warning flags at all to any galaxy (i.e.\ \texttt{drp3qual=0} for all galaxies).  We also require the DAP to have successfully modelled an emission spectrum cube for each galaxy.  These criteria produce a full sample of 6861 galaxies, of which 3255 belong to the Primary sample, 4342 to the Primary+ sample, and 2519 to the larger-coverage Secondary sample.

\subsection{Sample weightings}
\label{sec:Sample-weights}

As a result of MaNGA's sampling, none of the survey's galaxy samples are intrinsically volume-limited.  However, since the sampling strategies are well-defined, each galaxy can be assigned an appropriate weighting such that any analysis can be performed on an effectively volume-limited sample \citep{Wake+17}.  An implementation of these weightings is described in detail by \citet{Wake+17}, but here --- as in \citet{Peterken+20Morph} --- we use weightings generated using the method implemented by \citet{Sanchez+19}; see also \citet{RodriguezPuebla+20} and Calette et al.\ (in preparation) for further details.  This choice of which set of weightings to use was due to the \citet{Sanchez+19} weightings being more robust at lower stellar masses and more detailed in its treatment of galaxy colour.  However, since the sample-weighting calculations are similar, we find that the results shown here are unchanged when the \citet{Wake+17} sample weightings are used instead.  The adopted galaxy weights are reliable for galaxies with stellar mass $M_{\star}>10^{9}\,M_{\odot}$, making this the limit above which each of the properly-weighted samples are effectively volume-limited.



\subsection{Comparison to single-fibre spectra}
\label{sec:Sample-radii}

\begin{figure}
    \centering
    \includegraphics[width=0.8\columnwidth]{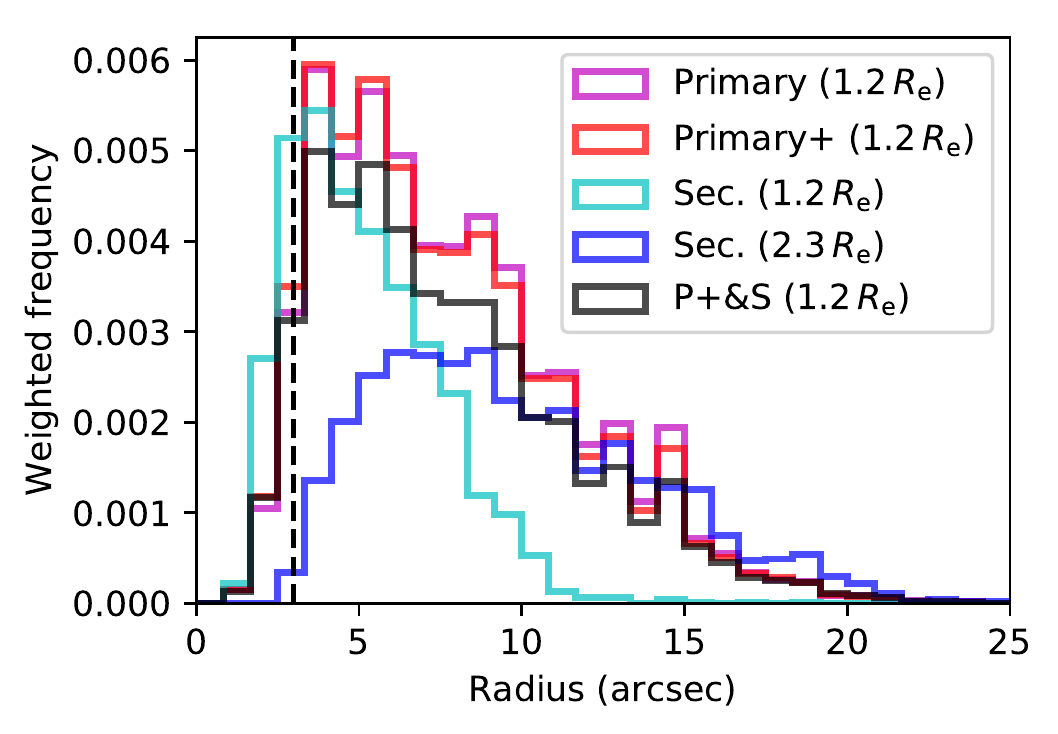}
    \caption{The weighted distributions of galaxy radii.  Each galaxy is sampled to $1.2\,R_{\rm e}$ (and $2.3\,R_{\rm e}$ for Secondary-sample galaxies), resulting in a distribution of on-sky apertures.  All samples probe to larger angular apparent radii than SDSS single-fibre spectra (marked as a vertical dashed line) would.}
    \label{fig:Radii}
\end{figure}

The weighted distributions of angular radii for each of the samples is shown in Figure~\ref{fig:Radii}.  For each of the samples, the majority of galaxies' $1.2\,R_{\rm e}$ radius threshold (as used in this work; see Section~\ref{sec:Fitting}) is larger than the $3\,\textrm{arcsec}$ radius probed by single-fibre SDSS spectra, highlighting the extra information available with integral-field spectroscopic observations.  As well as including more of the total star-formation in the Universe, the spectra we use here are measuring consistent radii in all galaxies regardless of their observed redshift.

\section{Spectral fitting}
\label{sec:Fitting}

In \citet{Peterken+20Morph}, we describe how we implemented full-spectrum fitting techniques to each Primary+ galaxy.  We use an identical method here but also including galaxies in the Secondary sample.  To summarise; we removed emission lines from each spaxel using the DAP's emission-line spectrum, and combined all spaxels' spectra within $1.2\,R_{\rm e}$ of each galaxy after removing line-of-sight velocities.  We then fit a single spectrum of the stellar component for each galaxy using \textsc{Starlight}, with a combination of 54 single stellar population (SSP) spectra from the E-MILES \citep{E-MILES} and \citet{Asa'd+17} libraries as templates.  We also assume a \citet{Calzetti+00} dust extinction model and fit within the range $3541.4\leq\lambda\leq8950.4\,\textrm{\AA}$.  Typical combined signal-to-noise ratios for each galaxy within the fitting range are at least $\sim500$.  Further description of the fitting method can be found in \citet{Peterken+20Morph}, and we also refer the reader to \citet{Peterken+20FR} for a full assessment of its reliability.

As well as fitting the spectrum of each galaxy sampled to $1.2\,R_{\rm e}$, we also repeat the above fitting method using all spaxels of all Secondary galaxies sampled to $2.3\,R_{\rm e}$.  We therefore obtained 9380 individual fits; one of each Primary+ galaxy, and two for each Secondary galaxy (sampled to each radius limit).  The two aperture limits were chosen to balance the inclusion of as much of the MaNGA field of view as possible, while avoiding overlap with the hexagonal IFU edges which might contaminate and bias results (see e.g.\ \citealt{IbarraMedel+16}).

We then obtain a $0.2\,\textrm{dex}$ time-smoothed star-formation history of each of the fits using the SSP weights assigned by \textsc{Starlight} using the method described in \citet{Peterken+20Morph}.  In smoothing, the \textsc{Starlight}-derived star-formation history (in units of $M_{\odot}\,\textrm{yr}^{-1}$) is first resampled at 250 lookback times which are evenly spaced in log(time) over the range of stellar populations used in fitting, and then convolved in the log(time) axis with a Gaussian function of width $0.2\,\textrm{dex}$.

\subsection{Comparison of measured star-formation rates}
\label{sec:Fitting-SFRsComparison}

\begin{figure}
    \centering
    \includegraphics[width=0.8\columnwidth]{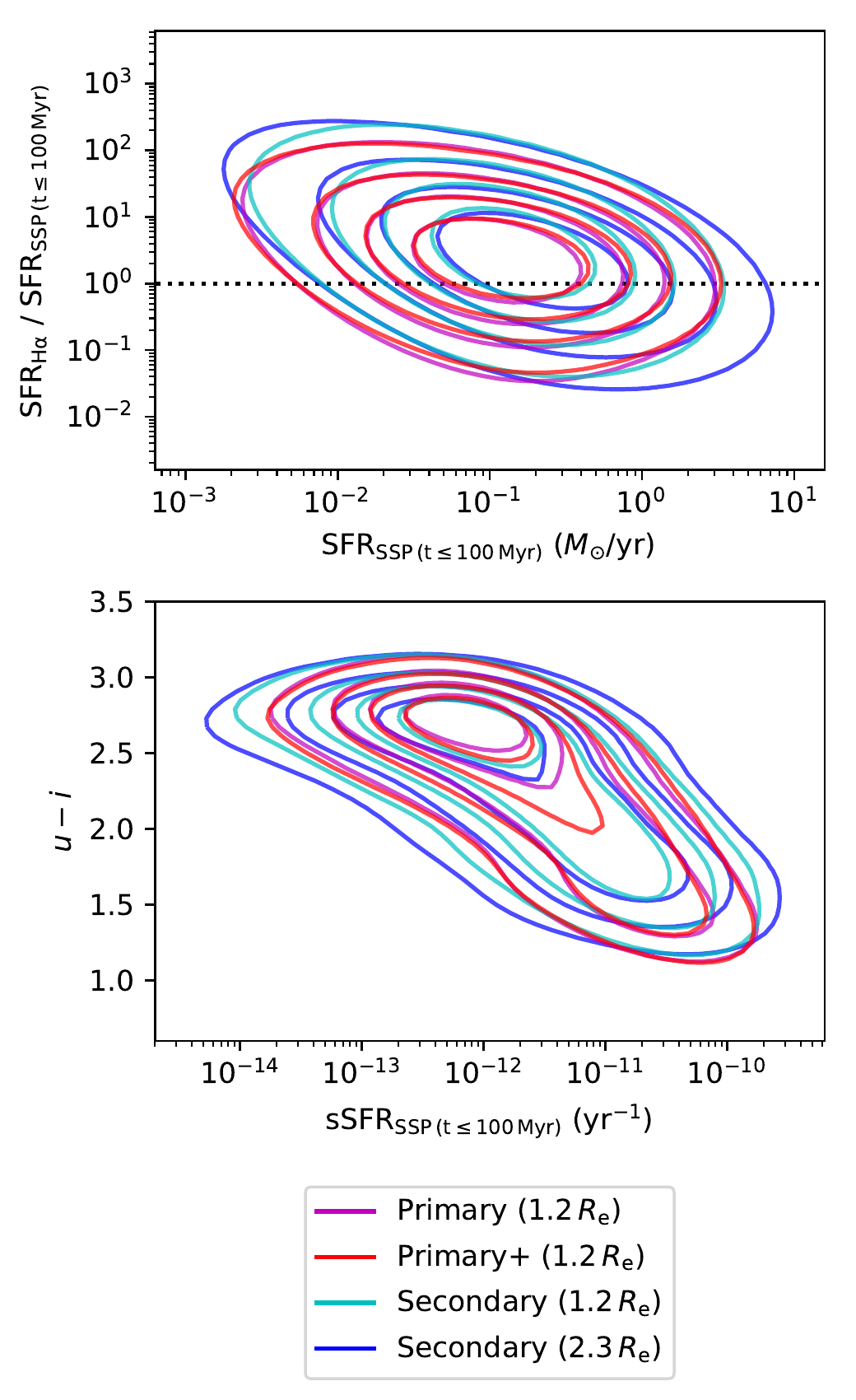}
    \caption{The present-day star-formation rates measured using both H$\alpha$ emission SFR$_{\rm H\alpha}$ are consistent with \textsc{Starlight}-derived total star-formation rate over the last $100\,\textrm{Myr}$ $\textrm{SFR}_{{\rm SSP},\, (t\leq100\,{\rm Myr})}$ (\textit{top}).  The NSA $u-i$ broad-band colours are also closely linked to the present-day specific star formation rate (\textit{bottom}).  Coloured contours indicate lines of 20\%, 40\%, 60\%, and 80\% of the peak density for each of the \textit{unweighted} samples.}
    \label{fig:SFR_Comparison}
\end{figure}

To establish the trustworthiness of the spectral fitting methods, we compare the measured outputs with known results obtained through established analyses.  We find that the total stellar mass measurements for each galaxy agree well with those determined photometrically by the NASA-Sloan Atlas (see \citealt{Peterken+20FR}).

We are also able to compare the present-day star-formation rates of each galaxy measured using the \textsc{Starlight}-derived star-formation histories with those calculated entirely independently from H$\alpha$ fluxes.  For each galaxy, we corrected the DAP's map of Gaussian-modelled H$\alpha$ emission line flux using the Balmer decrement, assuming an intrinsic value of $f_{\rm H\alpha}/f_{\rm H\beta}=2.87$ (corresponding to electron temperature $T_{\rm e}=10^{4}$~K and density $n_{\rm e}=10^{2}$~cm$^{-3}$ under \citet{OsterbrockFerland06} ``Case~B'' recombination) and a \citet{Calzetti+00} reddening curve with $R_V=3.1$.  Note that in the stellar population modelling in Section~\ref{sec:Fitting}, we assume $R_V=4.05$, but the lower value is used when considering the emission lines; see \citet{CatalanTorrecilla+15} and \citet{Greener+20} for an explanation on the difference between dust corrections applied to stellar and gas spectra.  These calculations and corrections are detailed fully by \citet{Greener+20}.

We then summed the flux from all spaxels within $1.2\,R_{\rm e}$ (and $2.3\,R_{\rm e}$ for the Secondary sample) which have an emission-line signal-to-noise ratio of at least 10 and calculated the star-formation rate $\textrm{SFR}_{\rm H\alpha}$ using the relation described by \citet{Kennicutt98}.  For consistency with the SSP templates used in fitting with \textsc{Starlight}, we assume a \citet{Chabrier03} IMF for this calculation.

We also calculate a present-day star-formation rate from \textsc{Starlight} $\textrm{SFR}_{{\rm SSP},\, (t\leq100\,{\rm Myr})}$ by extracting from the smoothed star-formation histories the total mass added to each galaxy over the last $100\,\textrm{Myr}$.  By plotting the offset between $\textrm{SFR}_{\rm H\alpha}$ and $\textrm{SFR}_{{\rm SSP},\, (t\leq100\,{\rm Myr})}$ as a function of $\textrm{SFR}_{{\rm SSP},\, (t\leq100\,{\rm Myr})}$, we show in Figure~\ref{fig:SFR_Comparison} that these two measurements of the ongoing star-formation rate are broadly comparable despite the H$\alpha$ emission having been removed from the spectra prior to fitting.  The offset from the line of equality is to be expected given that the two measurements are using completely different approaches and calibrations, contain differing systematics, and are sensitive to different timescales of star-formation.  For the Primary+ sample, we find that the relationship between the two measurements can be described by
\begin{equation}
    \label{eq:HaSSP}
    \log\left(\frac{\textrm{SFR}_{\rm H\alpha}}{\textrm{SFR}_{{\rm SSP},\, (t\leq100\,{\rm Myr}}}\right) = -0.5 \log\left(\textrm{SFR}_{{\rm SSP},\, (t\leq100\,{\rm Myr})}\right) -0.1
\end{equation}
with a root mean square deviation of $\sim0.7\,\textrm{dex}$ from this best-fit line.  Comparable relationships are found for the other samples.  The relatively small offset does not affect any conclusions, reassuring that the stellar population fits are reliable.

We obtain \textsc{Starlight}-derived specific star-formation rates $\textrm{sSFR}_{{\rm SSP},\, (t\leq100\,{\rm Myr})}$ by calculating the ratio of $\textrm{SFR}_{{\rm SSP},\, (t\leq100\,{\rm Myr})}$ to the total \textsc{Starlight}-measured stellar mass contained within each present day galaxy.  We also show in Figure~\ref{fig:SFR_Comparison} that such specific star-formation rates are closely correlated with a galaxy's $u-i$ NSA broadband colour, with a best-fit relation in the Primary+ sample described by
\begin{equation}
    \label{eq:ColoursSFR}
    u-i = -0.3 \log\left(\textrm{SFR}_{{\rm SSP},\, (t\leq100\,{\rm Myr})}\right) - 1.7
\end{equation}
which has a root mean square devation of $\sim0.4\,\textrm{mag}$.  As before, comparable relationships also exist for the other galaxy samples.  This link is unsurprising, as the redder $i$ band is most sensitive to the low-mass stars which comprise the bulk of a galaxy's total stellar mass, while the $u$ band is more sensitive to bluer stars and is therefore indicative of recent star-formation, making $u-i$ a suitable proxy for specific star-formation rate.

\section{The star-formation history of the Universe}
\label{sec:SFHoU}

Having obtained individual star-formation histories for each galaxy, the cosmic star-formation history can be constructed.  To do so, the individual star-formation histories measured for each galaxy must be carefully weighted and combined.  In combining galaxy star-formation histories, we account for the lookback time due to the observed reshift of each galaxy.  Each star-formation history's age sampling is shifted onto a lookback-time sampling by adding the galaxy's redshift's lookback time.  The galaxy star-formation histories are then combined and interpolated onto a common sampling of lookback times.


\begin{figure}
    \centering
    \includegraphics[width=0.78\columnwidth]{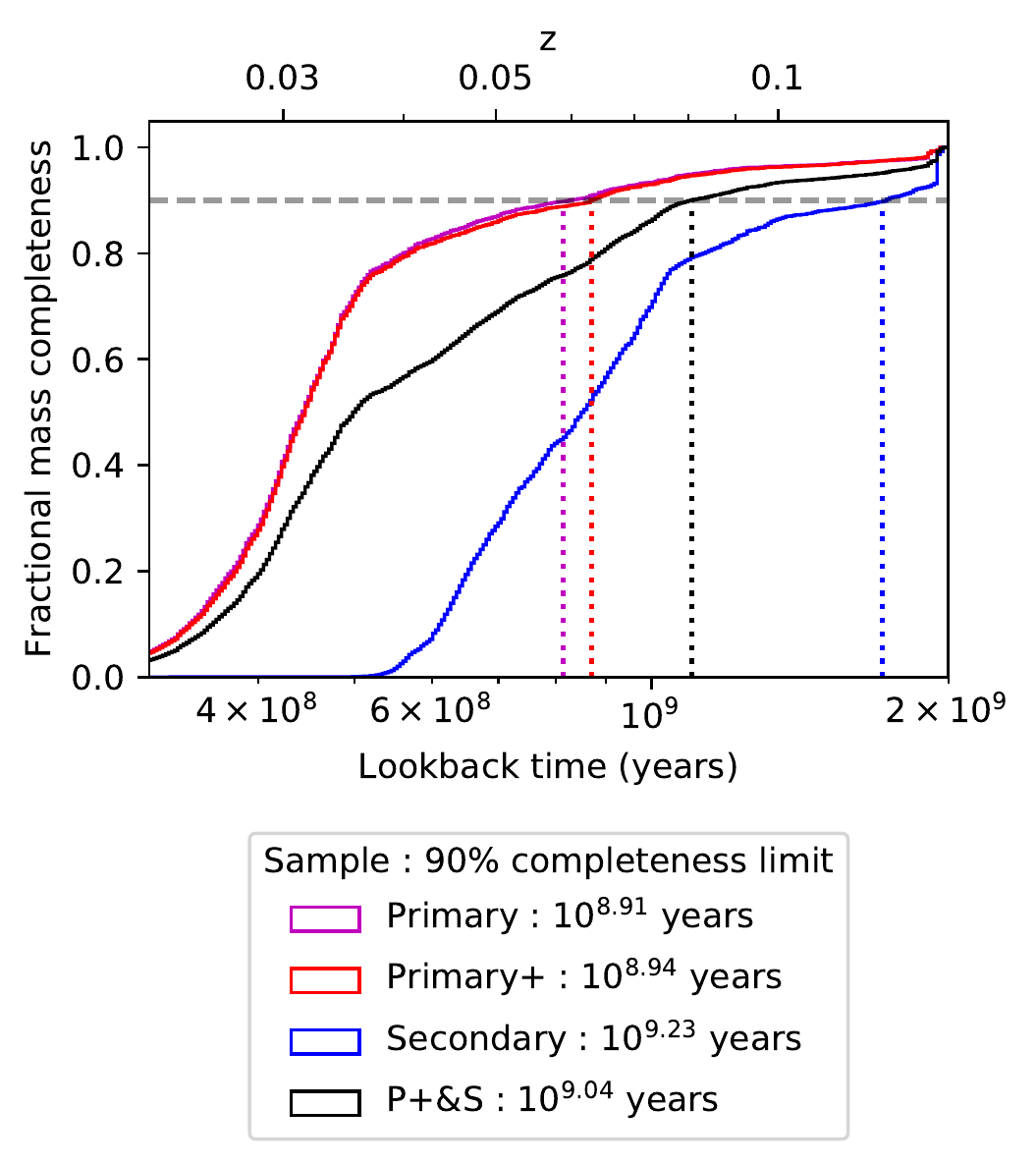}
    \caption{Fractional cumulative completeness in sample-weighted stellar mass as a function of lookback time, due to galaxies' observed redshifts.  The Secondary sample contains galaxies at higher redshifts than the Primary and Primary+ samples.  The lookback times corresponding to a 90\% mass completeness for each sample are given in the legend, and are illustrated by vertical dotted lines.  We do not sample star-formation histories at lookback times below these limits.}
    \label{fig:CompletenessLimits}
\end{figure}

Since the Primary(+) and Secondary samples are selected from different redshift distributions \citep{Wake+17} --- as shown in Figure~\ref{fig:CompletenessLimits} --- any derived star-formation history of the Universe will only be able to probe down to a specific limit in lookback times, depending on which sample is used.  Since the distribution of galaxy redshifts within any sample is dependent on the galaxy mass, we only measure cosmic star-formation histories at lookback times greater than that for which each sample has at least 90\% completeness in sample-weighted mass.  The specific adopted limits are shown in the legend in Figure~\ref{fig:CompletenessLimits}.

\begin{figure}
    \centering
    \includegraphics[width=0.8\columnwidth]{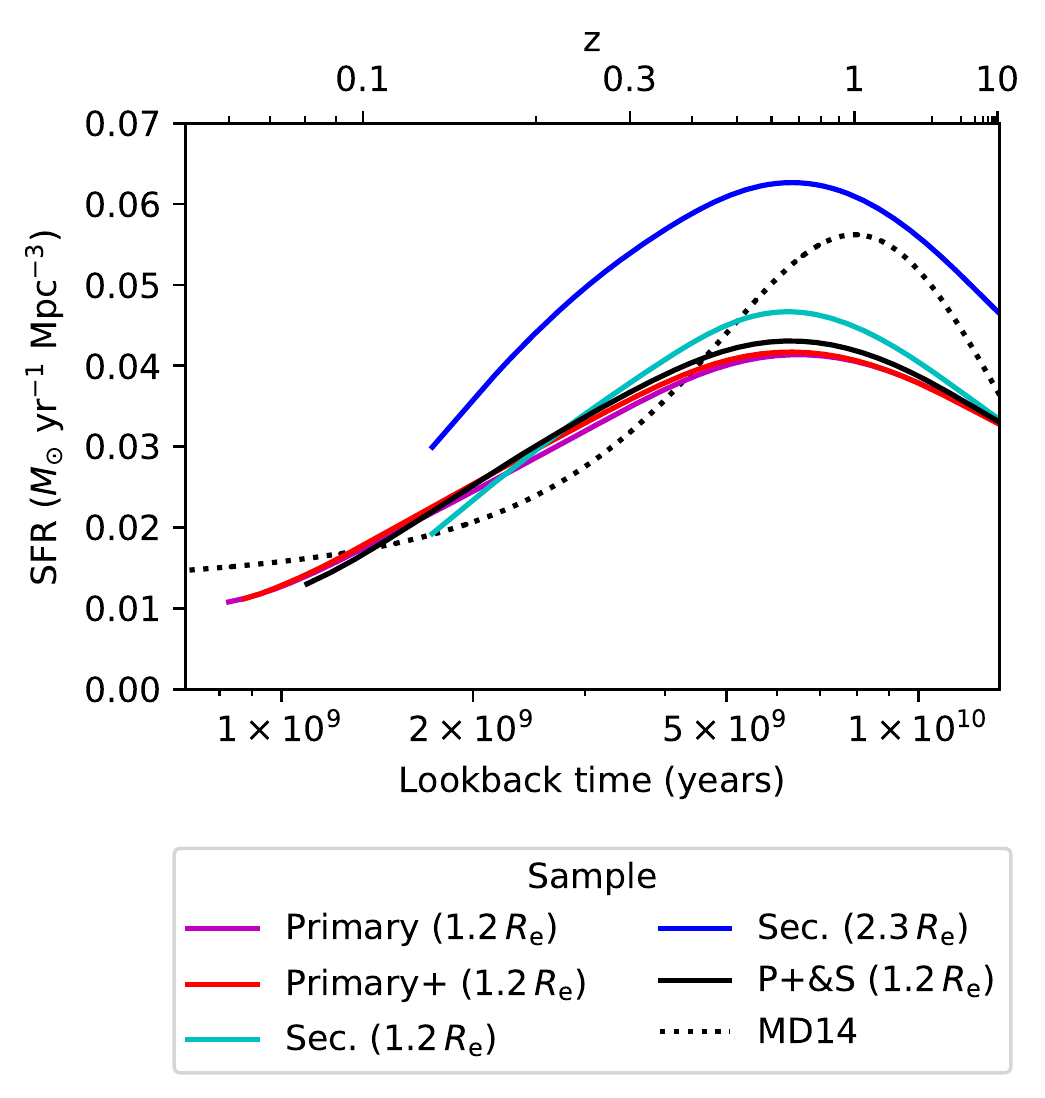}
    \caption{The star-formation history of the Universe determined using the different galaxy samples (solid coloured lines).  The agreement between the Primary and Primary+ samples is too close to be able to properly distinguish between them.  Also shown is the \citet{MD14} best-fit parametric function to the cosmic star-formation history derived from galaxy redshift studies (black dotted line).  For consistency, this comparison study has been smoothed according to our treatment of each individual galaxy's star-formation history.}
    \label{fig:SampleComparison}
\end{figure}

The cosmic star-formation histories measured from each sample is shown for each sample in Figure~\ref{fig:SampleComparison}.  We show that the calculated star-formation history of the Universe measured within 1.2~$R_{\rm e}$ of each galaxy agrees well at all lookback times regardless of the MaNGA sample used, and that all samples show a peak cosmic star-formation history at a lookback time of $10^{9.80\pm 0.01}\,\textrm{years}$ (corresponding to $z\approx0.67\pm0.02$).  

Figure~\ref{fig:SampleComparison} also shows the best-fit function derived by \citet{MD14} of the cosmic star-formation history measured through galaxy redshift studies.  To ensure a like-for-like comparison, we calculated the mass weights which would be assigned to each SSP template age used in our fitting method based on Equation~15 of \citet{MD14}, by integrating the curve within a box centred in log-space on each SSP's nominal age, to obtain a raw star-formation history as might ideally be measured using the stellar population fitting methods described in Section~\ref{sec:Fitting}.  We then smoothed the modelled SSP weights using the procedure described in Section~\ref{sec:Fitting} and in \citet{Peterken+20Morph}.  This smoothing procedure results in the smoothed profile showing a broader peak in the cosmic star-formation history which occurs at a lower redshift (at $z\sim1$ rather than $z\sim2$) compared to the unsmoothed \citet{MD14} best-fit function (not shown).

We find that the smoothed \citet{MD14} star-formation history is in general agreement with the results obtained through the completely independent approach performed here --- with particularly strong agreement at lookback times less than $\sim6\,\textrm{Gyr}$ ($z\sim0.63$) --- lending further evidence that such a fossil~record analysis is trustworthy.  We suggest that the broader peak obtained here is due to having co-added multiple star-formation histories which have each been smoothed to $0.2\,\textrm{dex}$ rather than simply smoothing a single function by $0.2\,\textrm{dex}$ in the comparison measurement.  We argue that this effect could also partly explain the difference in lookback times to the peak in cosmic star formation, although the observed difference is small given the entirely independent approaches of the two methods.  Our results also show good quantitative agreement with the cosmic star-formation histories obtained through fossil-record analyses performed by both \citet{LopezFernandez+18} and \citet{Sanchez+19}, and also with the earlier best-fit parametric models to observational redshift studies determined by \citet{Behroozi+13} and \citet{HopkinsBeacom06}.  

If the measured lookback time corresponding to the peak of cosmic star-formation was dictated primarily by artefacts of the SSP templates or of the fitting method, we would expect to find that the peak in each galaxy's individual star-formation history might be biased towards a certain stellar population age.  In such a case, the derived cosmic star-formation history measured in each sample would therefore display its peak at \textit{different} lookback times once galaxies' observed redshifts are accounted for, since each MaNGA sample targets galaxies from different redshift distributions.  The close match between star-formation histories shown in Figure~\ref{fig:SampleComparison} therefore shows that the signal being measured is intrinsic to the data rather than being the product of artefacts.  Indeed, we find that the agreement between the star-formation histories measured using each MaNGA sample is \textit{only} seen when the galaxy redshifts are taken into account, and that the above close agreement is smaller than the difference in median lookback~times for each sample's galaxy populations.

\medskip

We find that increasing the aperture from $1.2\,R_{\rm e}$ to $2.3\,R_{\rm e}$ using the Secondary sample results in a cosmic star-formation history which is greater at all lookback times than \citet{MD14}'s best fit to cosmological results.  This excess could either be due to overly-conservative aperture corrections or to surface brightness limits causing cosmological studies to have underestimated the total star-formation rates at different redshifts.

\begin{figure}
    \centering
    \includegraphics[width=0.78\columnwidth]{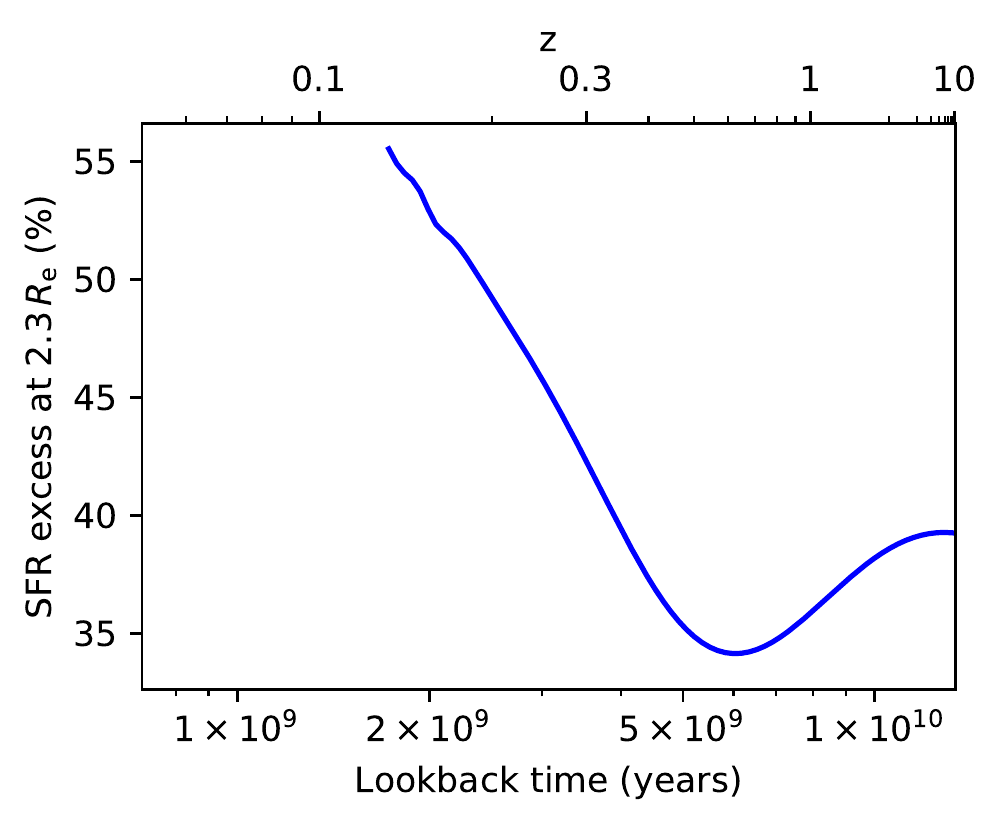}
    \caption{The fractional excess in measured cosmic star formation at all lookback times from the Secondary sample when the aperture is increased from $1.2\,R_{\rm e}$ to $2.3\,R_{\rm e}$.  At more recent times, the fractional increase is larger, indicative of inside-out growth occurring in most galaxies.}
    \label{fig:SFHExcess}
\end{figure}

Having derived the cosmic star-formation history using the Secondary-sample galaxies with radius limits of $1.2\,R_{\rm e}$ and $2.3\,R_{\rm e}$ separately, we are able to quantify what extra fraction of star-formation is included by increasing the FOV diameter by a factor of $\sim1.9$.  Comparison between the two Secondary-derived cosmic star-formation histories of Figure~\ref{fig:SampleComparison} shows that the effect of the increased FOV results in a 35--40\% enhancement in the measured star-formation histories at lookback times $\gtrapprox 5\,\textrm{Gyr}$, increasing to a 55\% enhancement by $\sim 1.5\,\textrm{Gyr}$, as illustrated in Figure~\ref{fig:SFHExcess}.  This increase over time is indicative of inside-out growth resulting in a greater star formation contribution by the galaxy outskirts at more recent times, as we explored for spiral galaxies in \citet{Peterken+20FR}; see also \citet{Perez+13, IbarraMedel+16, GarciaBenito+17, Goddard+17}.  It is interesting to note that even at recent times, the ratio in star-formation rate between larger to the smaller aperture of $\sim1.5$ is smaller than the corresponding ratio of sky coverage area ($\sim3.7$), showing that star-formation density is still greatest at galactic centres on average despite the effect of inside-out growth.

\subsection[Size: effects of present-day stellar mass]{Size\footnote{We are not using ``size'' here in its usual meaning of a galaxy's physical or apparent radius or diameter; rather, we mean its stellar mass.  Still, such a simple substitution is suitable to sufficiently sustain a satisfactory sibilance with subsequent subheadings.}: effects of present-day stellar mass}
\label{sec:SFHoU-Mass}

\begin{figure}
    \centering
    \includegraphics[width=0.8\columnwidth]{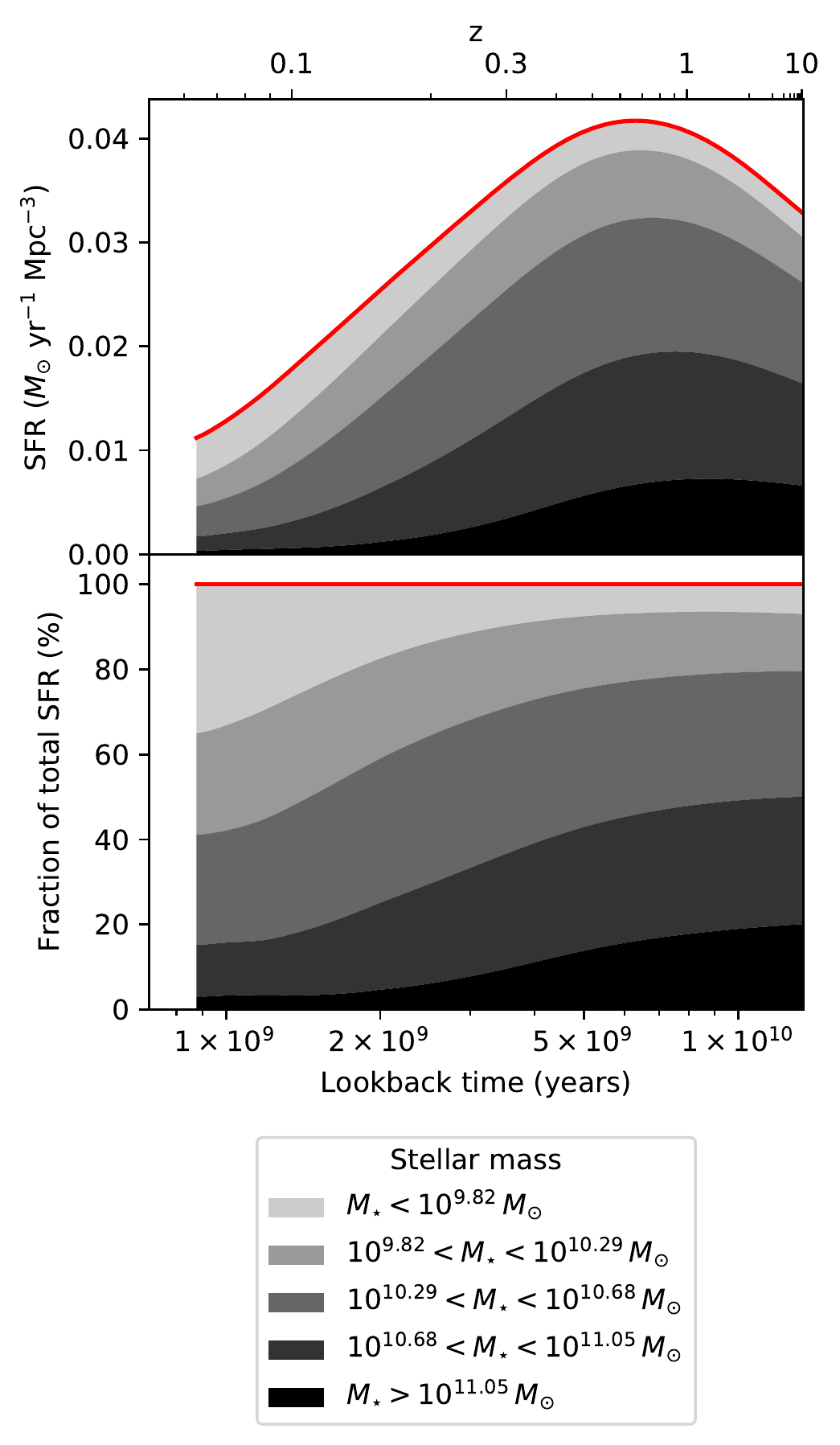}
    \caption{The star-formation history of the Universe stratified by its contributions from galaxies in different present-day stellar mass bins, in absolute star-formation rates (\textit{top}) and by the fractional contribution from each mass bin (\textit{bottom}).  Shown is for the Primary+ sample (red line), but all other samples are similar.  Higher-mass galaxies have become less dominant in the more recent Universe and vice-versa.}
    \label{fig:Masses}
\end{figure}

Having obtained the cosmic star-formation history, we now begin to explore the connection between a galaxy's present-day physical properties with its historical contribution to the Universe's star-formation.  We split the galaxy sample into five discrete bins of present-day photometrically-measured stellar mass $M_{\star\,{\rm NSA}}$ such that each bin contains an equal (unweighted) number of Primary+-sample galaxies.  The thresholds of these bins and the star-formation history of the Universe split into contributions of these mass bins are shown in Figure~\ref{fig:Masses}.  We have shown the stellar-mass breakdown for the Primary+ sample within $1.2\,R_{\rm e}$, but all other samples' are similar.  We see that the star-formation contribution from higher-mass galaxies becomes less significant at more recent times.  For example, galaxies with present-day stellar mass $M_{\star}<10^{10.29}\,M_{\odot}$ dominated the cosmic star formation until $\sim2\,\textrm{Gyr}$ ago to contribute only $\sim50\%$ in the local Universe.  We have therefore recovered the known observational effects of downsizing.

\begin{figure}
    \centering
    \includegraphics[width=0.9\columnwidth]{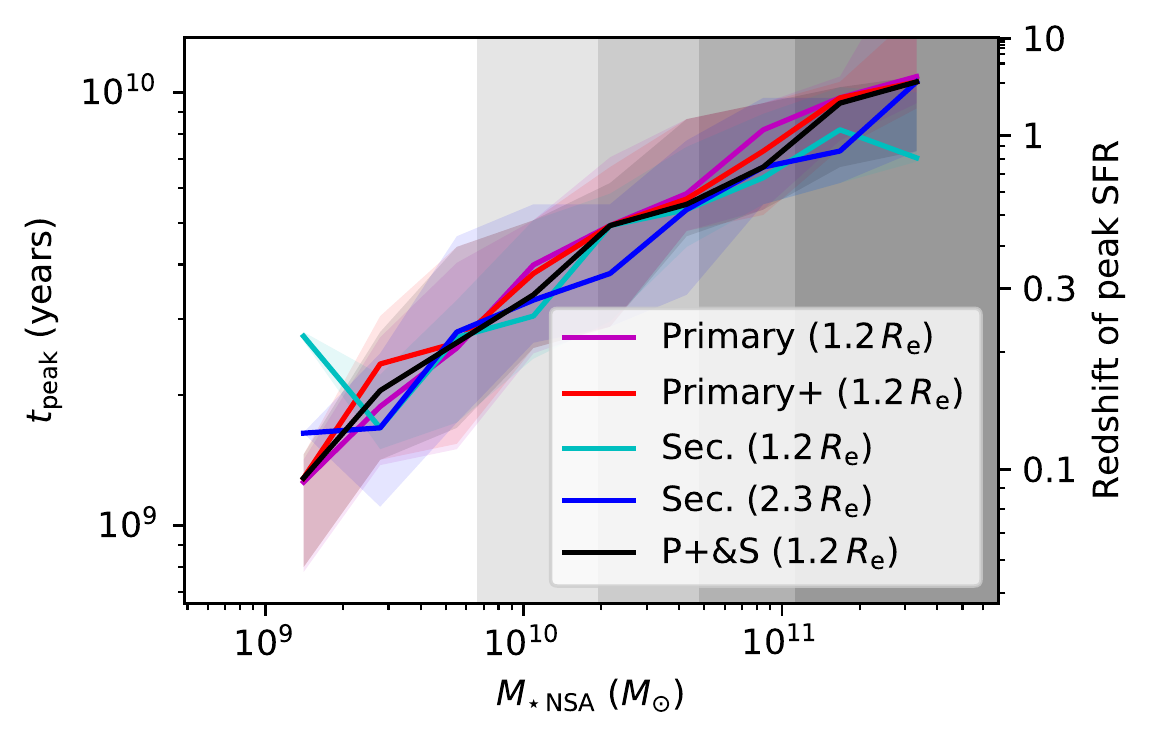}
    \caption{The lookback time to the peak of star-formation $t_{\rm peak}$ as a function of present-day stellar mass $M_{\star\,{\rm NSA}}$, using each of the MaNGA samples.  The solid lines indicate the sample-weighted median values, and the shaded regions lie between the one-third and two-thirds percentiles.  Greyscale-shaded background regions indicate the mass bins used in Figure~\ref{fig:Masses} and elsewhere.}
    \label{fig:PeaksBySample}
\end{figure}

A similar perspective of downsizing is gained by measuring the lookback time at which each individual galaxy reached its maximum star-formation rate.  The volume-weighted median lookback time $t_{\rm peak}$ of peak star-formation in galaxies as a function of their present-day stellar mass $M_{\star}$ is shown in Figure~\ref{fig:PeaksBySample}.  We find that the peak in star formation typically occurred at $t_{\rm peak} = 8\pm1\,\textrm{Gyr}$ ago (corresponding to $z\approx1.1\pm0.3$) in the galaxies with the highest present-day stellar masses in the samples ($M_{\star}>10^{11}\,M_{\odot}$), falling to $t_{\rm peak}\lessapprox 2\,\textrm{Gyr}$ ($z\lessapprox0.2$) for galaxies with present-day stellar masses of $M_{\star}<3\times10^{9}\,M_{\odot}$.  We also see some signs of downsizing effects being strongest among low-mass galaxies where the gradient $\Delta t_{\rm peak}/\Delta M_{\star}$ is greatest.

By measuring the lookback time by which galaxies had built the bulk of their present-day stellar mass, \citet{IbarraMedel+16}, \citet{GarciaBenito+17}, and \citet{Peterken+20FR} have previously found that low-mass galaxies typically show larger variation in the characteristic formation times than high-mass galaxies.  However, we find here that the scatter in $t_{\rm peak}$ remains approximately constant at $\sim0.3\,\textrm{dex}$ over all stellar masses.  The two results are not incompatible: we interpret such an apparent dichotomy as indicative of low-mass galaxies having greater variation in the rate of decline in star-formation after their common peak time.

Figure~\ref{fig:PeaksBySample} again shows strong agreement between each MaNGA sample.  We therefore present hereafter only results measured using the Primary+ sample, but results for other samples are similar throughout.

\subsection{Shade: effects of present-day colour}
\label{sec:SFHoU-Col}

By splitting each galaxy sample by the galaxies' locations on the stellar mass--colour plot, we are also able to investigate the effect of present-day colour on a galaxy's contribution to the cosmic star-formation history.  

\subsubsection{Colour classifications}
\label{sec:SFHoU-Col-Classes}

\begin{figure}
    \centering
    \includegraphics[width=\columnwidth]{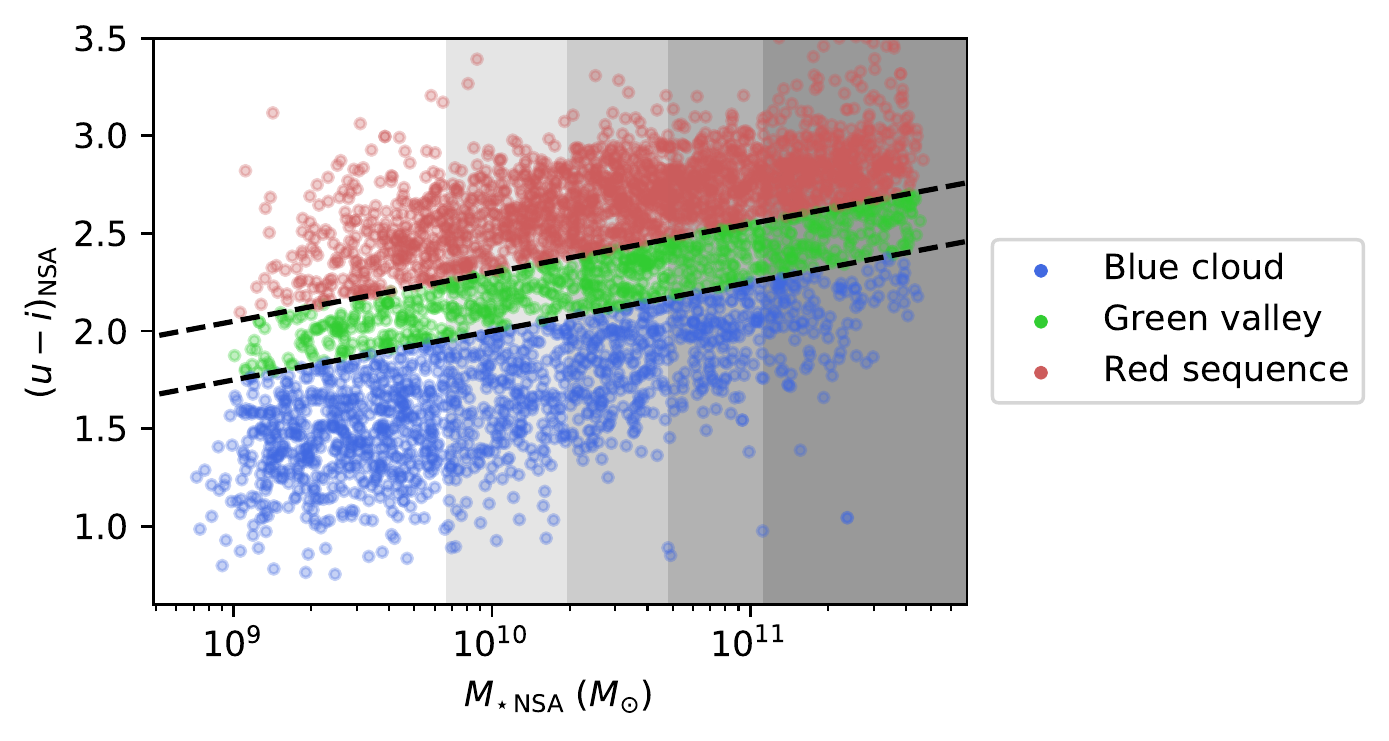}
    \caption{Mass--colour plane of the Primary+-sample galaxies, with galaxies classified according tho their colour as used in Section~\ref{sec:SFHoU-Col}.  Galaxy stellar masses $M_{\star\,\textrm{NSA}}$ and $\left(u-i\right)_{\textrm{NSA}}$ broadband colours are NSA measurements.  Shaded backgrounds indicate the bins in stellar mass in Figure~\ref{fig:Masses} and elsewhere.}
    \label{fig:col-mass_col}
\end{figure}

We separate galaxies by their colours according to the following criteria:
\begin{itemize}
    \item Red sequence: $\left(u-i\right)_{\rm NSA} > 0.25\log\left(M_{\star\,{\rm NSA}}\right)-0.2$
    \item Green valley:\\ $0.25\log\left(M_{\star\,{\rm NSA}}\right)-0.5 \leq \left(u-i\right)_{\rm NSA} \leq 0.25\log\left(M_{\star\,{\rm NSA}}\right)-0.2$
    \item Blue cloud: $\left(u-i\right)_{\rm NSA} < 0.25\log\left(M_{\star\,{\rm NSA}}\right)-0.5$
\end{itemize}
where the galaxies' stellar masses $M_{\star\,{\rm NSA}}$ (measured in $M_{\odot}$) and $\left(u-i\right)_{\rm NSA}$ broadband colours are taken from the NSA \citep{NSA} photometry-derived measurements.\footnote{Note that a galaxy's broadband colour is affected by its inclination due to increased internal dust extinction.  We do \textit{not} attempt to correct for this effect, so that our measurements resemble those of other studies of galaxy populations as closely as possible.}  These thresholds and the subsequent classifications of Primary+ galaxies are illustrated in Figure~\ref{fig:col-mass_col}.

\begin{figure}
    \centering
    \includegraphics[width=\columnwidth]{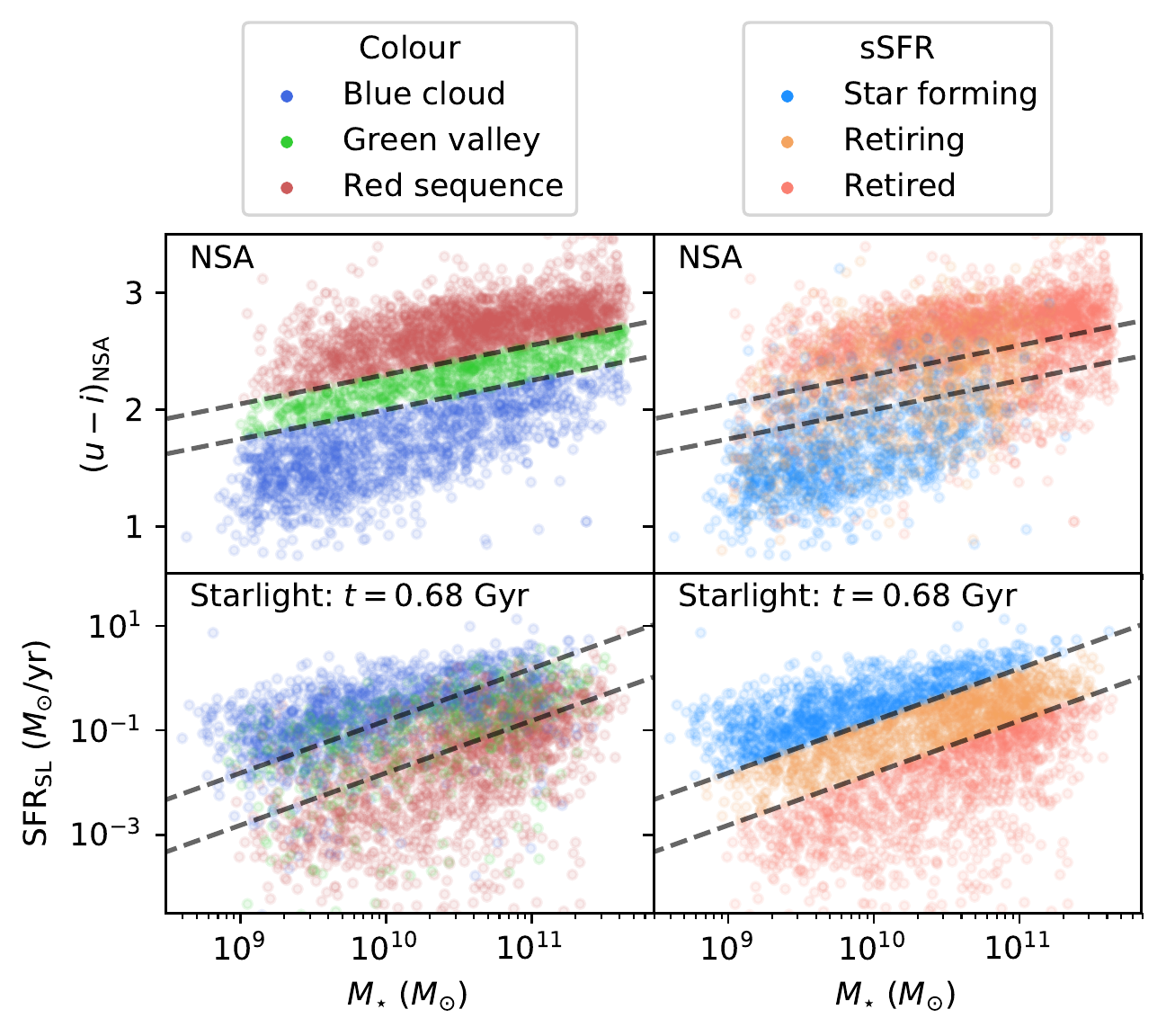}
    \caption{Comparison of galaxies with colour classifications defined in Section~\ref{sec:SFHoU-Col} (\textit{left panels}, based on $\left(u-i\right)_{\rm NSA}$ broadband colours) and the specific star-formation rate classifications defined in \citet{Peterken+20Morph} (\textit{right panels}) in the $\left(u-i\right)_{\rm NSA}$--$M_{\star\,{\rm NSA}}$ plane (\textit{top panels}) and $\textrm{SFR}_{\rm SL}$--$M_{\star\,{\rm SL}}$ (\textit{bottom panels}) plane at a lookback time of $t=0.68\,\textrm{Gyr}$ for Primary+ galaxies.  The classifications of blue~cloud, green~valley, and red~sequence are analogous to classifications of star-forming, retiring, and retired galaxies respectively.  Dashed lines in the \textit{top panels} indicate the boundaries used to separate blue cloud, green valley, and red sequence galaxies, and those in the \textit{bottom panels} indicate those used to separate star-forming, retiring, and retired galaxies.}
    \label{fig:Colour_SFR_Comparison}
\end{figure}

We showed in Section~\ref{sec:Fitting-SFRsComparison} and Equation~\ref{eq:ColoursSFR} that broadband colour and specific star-formation rate are closely correlated.  To demonstrate how the above colour classifications are linked to star-formation properties, we have also split the galaxy sample into star-forming, retiring, and retired populations using thresholds of $\textrm{sSFR}_{{\rm SL}\left(t=0.68\,{\rm Gyr}\right)} = 10^{-11}$ and $10^{-12}\,\textrm{yr}^{-1}$ --- as used in \citet{Peterken+20Morph} --- where $\textrm{sSFR}_{{\rm SL}\left(t=0.68\,{\rm Gyr}\right)}$ is the ratio of the average \textsc{Starlight}-measured instantaneous star-formation rate to the \textsc{Starlight}-measured instantaneous stellar mass $M_{\star\,{\rm SL}}$ at a lookback time of $t=0.68\,{\rm Gyr}$, which is the lowest lookback time measurable with the Primary+ sample as defined in Figure~\ref{fig:CompletenessLimits}.  Figure~\ref{fig:Colour_SFR_Comparison} shows a direct comparison of galaxies' classifications under both schemes on the $\left(u-i\right)_{\rm NSA}$--$M_{\star\,{\rm NSA}}$ plane (using NSA-measured colours $\left(u-i\right)_{\rm NSA}$ and stellar masses $M_{\star\,{\rm NSA}}$) and on the $\textrm{SFR}_{\rm SL}$--$M_{\star\,{\rm SL}}$ plane at a lookback time of $t=0.68\,\textrm{Gyr}$ (using \textsc{Starlight}-derived star-formation rates SFR$_{\rm SL}$ and stellar masses $M_{\star\,{\rm SL}}$).  We find that most star-forming galaxies map onto the blue cloud region, most retiring galaxies to the green valley region, and most retired galaxies to the red sequence region, and vice-versa, showing that the $u-i$ colour classifications defined above are broadly analogous to classifications of star formation\footnote{This strong link is by constuction; we chose the $u-i$ broadband colour in Section~\ref{sec:Fitting-SFRsComparison} as a proxy for specific star-formation rate due to the sensitivity of $u$ to recent star-formation and $i$ to the total stellar mass.  Classifications based on broadband colours at longer wavelengths (e.g.\ $i-z$) will not be so closely linked to the classifications of specific star-formation rates.}, and therefore linking the observed measurements to physical galaxy properties.

Having demonstrated the commutativity of broadband colour and specific star-formation rate, it would be feasible to investigate how either of these properties relates to a galaxy's history.  \citet{IbarraMedel+16} showed that colour and specific star-formation rate have similar effects on the star-formation history of galaxies.  However, since colour is a more readily-measurable property of a galaxy, we will continue to investigate the role of present-day broadband colour rather than present-day specific star-formation rate in the analysis shown here, for more straightforward application of the results to other studies of galaxy populations.

\subsubsection{Present-day colour and the star-formation history of the Universe}
\label{sec:SFHoU-Col-Effect}

\begin{figure}
    \centering
    \includegraphics[width=0.8\columnwidth]{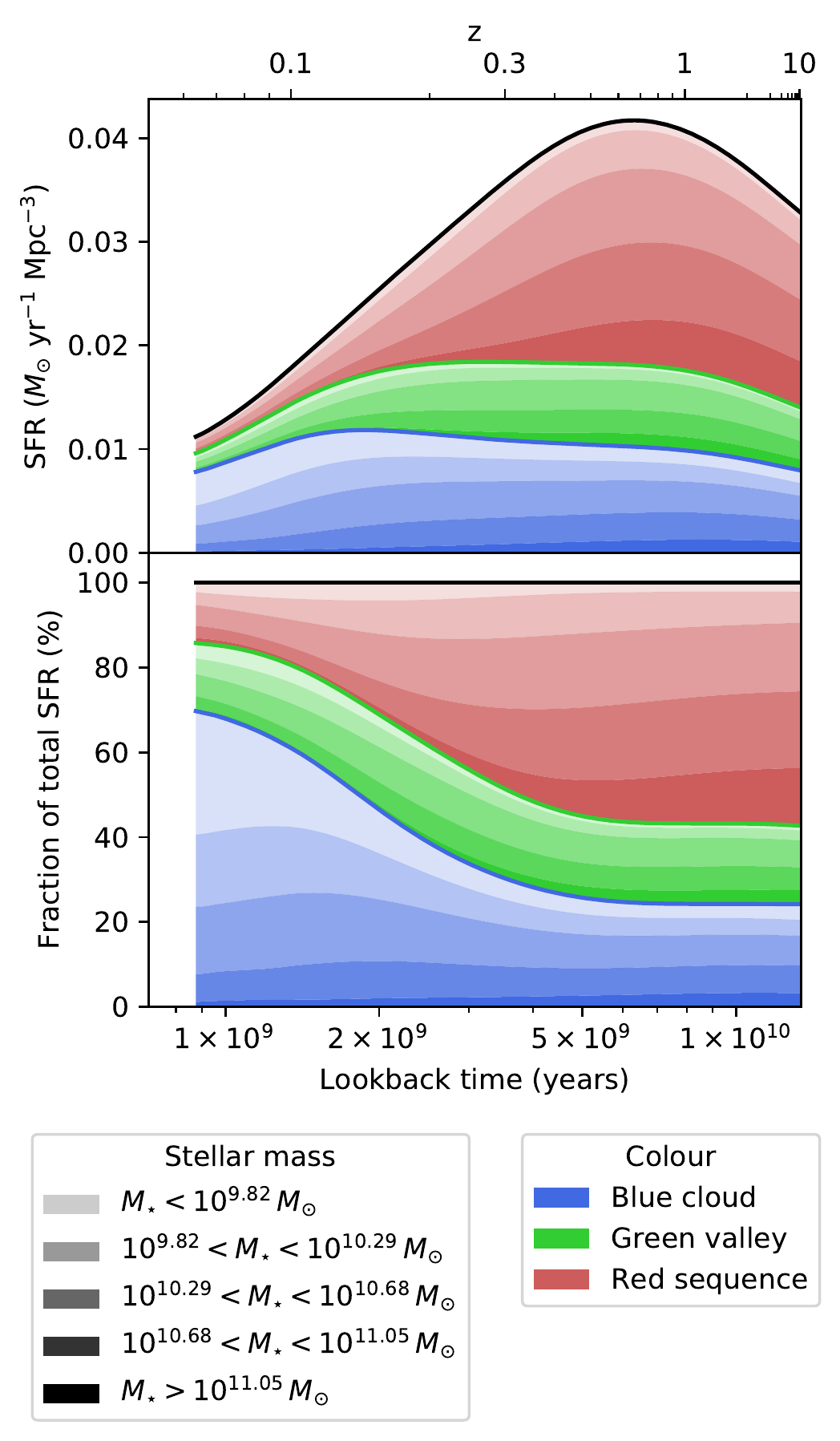}
    \caption{The star-formation history of the Universe inferred from the Primary+ sample, separated into its contributions from galaxies belonging to each galaxy classification (indicated by colours) defined in Section~\ref{sec:SFHoU-Col-Classes}.  Each colour classification is also stratified by present-day stellar mass (darkest to lightest shades for most to least massive galaxies) within each broadband colour classification, using the mass bins of Section~\ref{sec:SFHoU-Mass}.  The \textit{upper panel} shows the total contribution to the star-formation history, while the \textit{lower panel} shows the fractional contributions to the total star-formation history.}
    \label{fig:Morphs_col}
\end{figure}

Figure~\ref{fig:Morphs_col} shows the Primary+-derived star-formation history of the Universe showing the relative contributions from galaxies classified as currently belonging to the blue cloud, the green valley, and the red sequence as defined above.  There is a strong correlation between a galaxy's present-day colour and its past contribution to the total star formation, with the blue cloud contributing only 20\% at early times and rising to more than 70\% at the present day.  We recover how galaxies which are currently exhibiting very low levels of star formation were once the dominant source of new stars.

\begin{figure*}
    \centering
    \includegraphics[width=\textwidth]{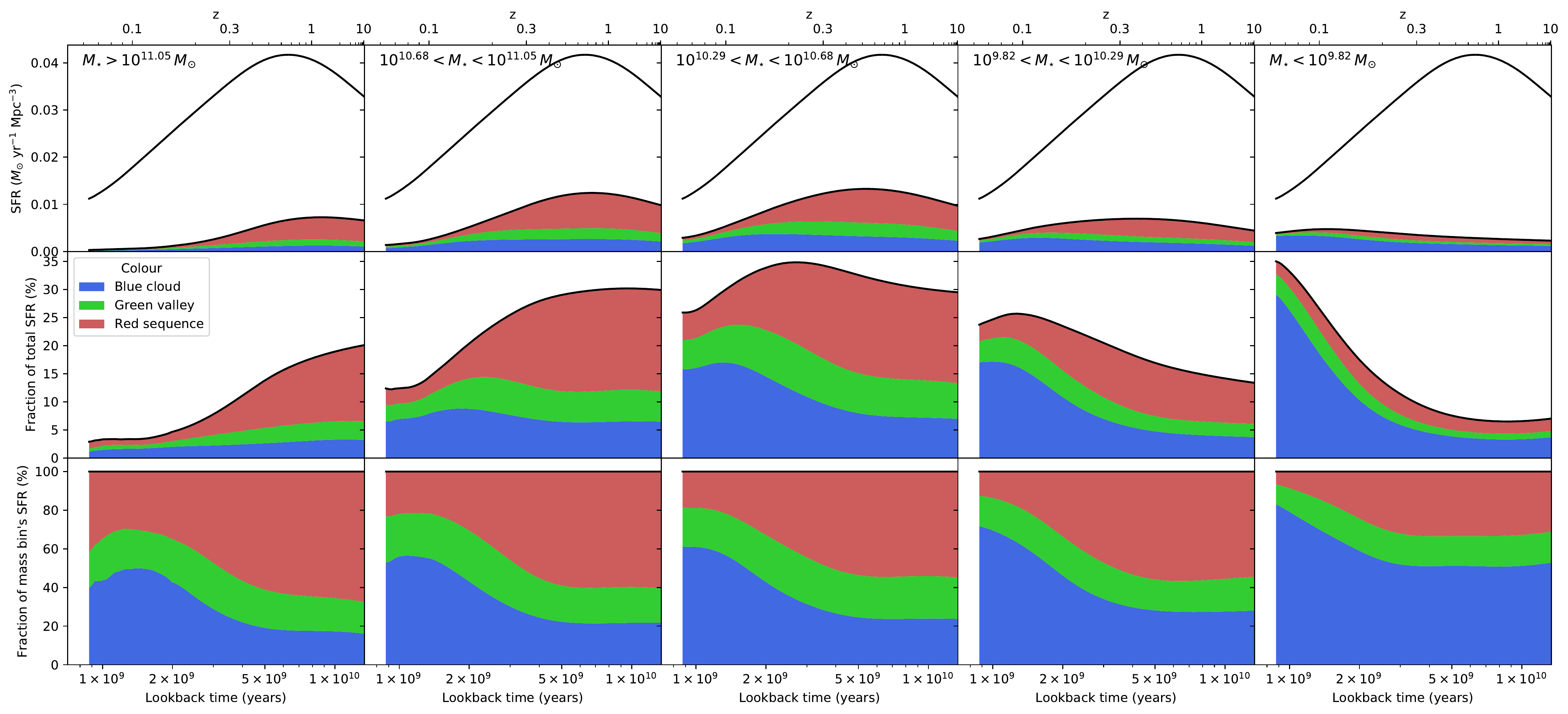}
    \caption{The absolute (\textit{top row}) and fractional (\textit{middle row}) contributions to the star-formation history of the Universe of galaxies of different present-day colour classifications (\textit{colours}) in each mass bin (\textit{columns}; transparencies equivalent to those used in Figure~\ref{fig:Morphs_col}), and the relative contribution of each colour classification to each mass bin's star-formation history (\textit{bottom row}).}
    \label{fig:MassSplits_col}
\end{figure*}

However, stellar mass and colour classification are not independent galaxy properties, as can be seen in Figure~\ref{fig:col-mass_col}.  Specifically, high-mass galaxies are more likely to belong to the red~sequence and vice-versa.  To determine how the effect of present-day colour varies with a galaxy's stellar mass, we show the relative contribution of each colour classification to each mass bin's star-formation history in Figure~\ref{fig:MassSplits_col}.  We find that the blue cloud's increase in contribution to the star-formation history between lookback times of 10 and 1~Gyr varies from a $\sim20$\% increase in the highest-mass bin to a $\sim40$\% increase in the lowest-mass bins:\footnote{The discrepancy between these mass-bin values and the increase of $\sim50$\% in the blue cloud's contribution overall is due to the dependent nature of colour and stellar mass: lower-mass bins with a larger fraction of blue cloud galaxies becoming more dominant at more recent times.} a galaxy's colour is therefore more strongly correlated to its historical contribution to the star-formation history in low-mass than in high-mass galaxies.

We additionally see some evidence for the colour designations reflecting shorter star-formation timescales in low-mass galaxies than in their high-mass counterparts.  The most massive blue cloud galaxies began increasing their contribution to that mass bin's star-formation history $\sim6\,\textrm{Gyr}$ ago to reach its current level $\sim2\,\textrm{Gyr}$ ago, while the lowest-mass bin's blue cloud only started to become more dominant $\sim2\,\textrm{Gyr}$ ago and appears to still be increasing to the present-day.

We find that galaxies currently in the green valley have contributed an approximately constant 20\% to the star-formation history in all mass bins.  Such a consistency of the green valley's contribution between mass bins is to be expected assuming that the timescale of star-formation transition is independent of stellar mass.  However, the constancy of the present-day green valley's contribution over the last $10\,\textrm{Gyr}$ indicates a complex picture.  If these galaxies are those caught in the act of a rapid ``quenching'' transition from the blue cloud to the red sequence or a rapid rejuvenation transition in the opposite direction, it would be expected that their historical contributions to the cosmic star-formation history should be similar to either the blue cloud's or the red sequence's until recent times.  The fact that this does not hold true could hint towards the possibility that the galaxies designated as belonging to the green valley using the colour criteria above contain a mix of quenching and rejuvenating galaxies.  However, further analysis of the individual star-formation histories of galaxies currently in the green valley shows that this is not true in that most show declining levels of star-formation over the last Gyr, and there is no significant population of rejuvenating green valley galaxies.  Instead, it seems that most present-day green valley galaxies have always had low but sustained levels of star formation, with star-formation histories which have generally followed the star-formation history of the Universe as a whole.

While we do not speculate here as to a possible physical explanation for this observation, it might reflect the slower retiring processes proposed by \citet{Schawinski+14}, for example.  Possible drivers of such a ``slow~quenching'' scenario are thought to be predominantly internal processes \citep{Martig+09, Fang+13, Bluck+14, Smethurst+18, Das+21}.  There is observational evidence to suggest that cessation of star-formation occurs on long (several Gyr) timescales in a significant number of galaxies, including ellipticals (see e.g.\ \citealt{Smethurst+15, Smethurst+17, Belfiore+18, Lacerna+20}), although other studies argue otherwise (see e.g.\ \citealt{Bremer+18}).

\begin{figure*}
    \centering
    \includegraphics[width=\textwidth]{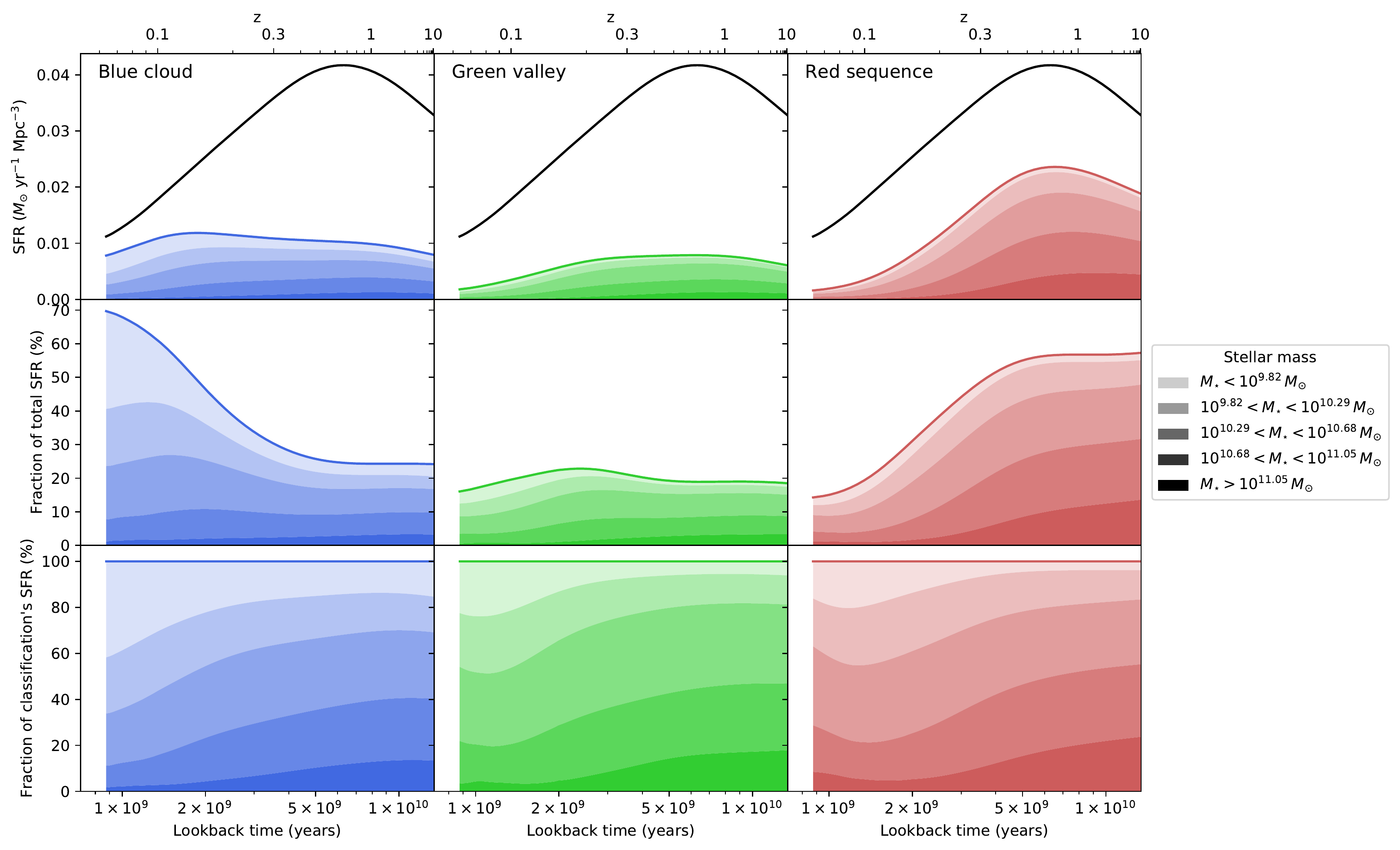}
    \caption{The absolute (\textit{top row}) and fractional (\textit{middle row}) contributions to the star-formation history of the Universe of galaxies of different stellar masses in each colour classification, and the relative contribution of galaxies with different stellar mass to each colour classification's star-formation history (\textit{bottom~row}).}
    \label{fig:MorphSplits_col}
\end{figure*}

By considering how galaxies of each colour classification in the five bins of stellar mass used in Section~\ref{sec:SFHoU-Mass} have contributed to the cosmic star-formation history, we see in Figure~\ref{fig:Morphs_col} some signs of downsizing effects being present in all colour classifications; low-mass galaxies of all colours become relatively more dominant at smaller lookback times.  To measure how downsizing effects vary between colour classifications in this way, we also show the relative contribution of galaxies in each mass bin to each colour classification's star-formation history in Figure~\ref{fig:MorphSplits_col}.  We find evidence for downsizing having occurred most strongly in galaxies currently in the blue cloud, with galaxies in the two lowest-mass bins increasing their contribution by 20\% in the red sequence and 30\% in the blue cloud, with the green valley lying in between.

\begin{figure}
    \centering
    \includegraphics[width=0.9\columnwidth]{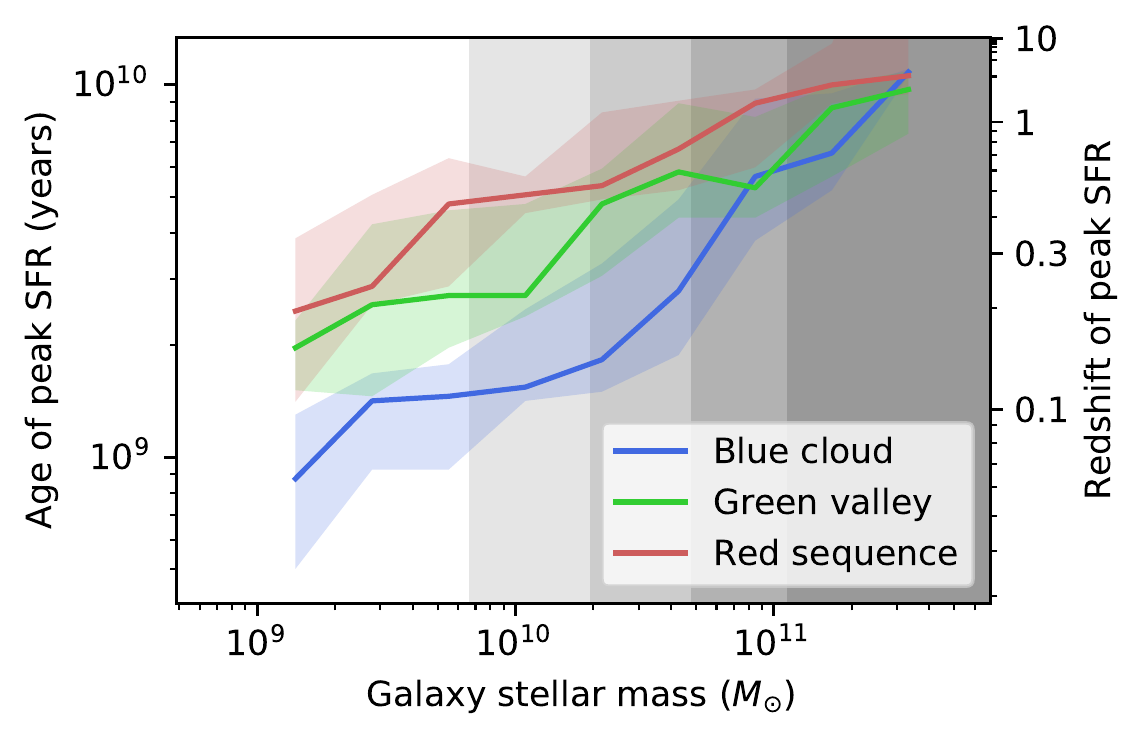}
    \caption{The lookback time to the peak star formation rate of galaxies currently belonging to each of the colour classifications.  Solid lines indicate median values, and shaded regions lie between the one- and two-third percentiles.  Blue cloud galaxies peaked in star formation more recently and show stronger downsizing effects than red sequence galaxies.}
    \label{fig:PeaksByColour}
\end{figure}

This difference in the strength of the downsizing effect seen in galaxies of different colour classifications is illustrated further in Figure~\ref{fig:PeaksByColour}, where we see that red sequence galaxies reached peak star formation at earlier times than green valley or blue cloud galaxies at all stellar masses, but that this effect is greatest in low-mass galaxies.

\subsection{Shape: effects of present-day morphology}
\label{sec:SFHoU-Morph}

Having quantitatively measured the link between a galaxy's present day colour and its historical contribution to the cosmic star-formation history, we can now independently assess the relative role of a galaxy's present-day morphology.

\subsubsection{Morphological classifications}
\label{sec:SFHoU-Morph-Classifications}

Over 91\% of galaxies in all MaNGA samples have been classified by the Galaxy Zoo ``citizen~science'' project \citep{Lintott+08}, in which volunteers are asked to classify galaxies' morphological features.  We make use of the redshift-debiased vote fractions of \citet{Hart+16} from Galaxy~Zoo~2 \citep{Willett+13} to split the galaxy samples by their present-day morphology.  We classify each galaxy according to the following criteria:
\begin{itemize}
    \item Elliptical: $p_{\rm features\, or\, disk}<0.5$
    \item S0: $p_{\rm features\, or\, disk}>0.5$ and $p_{\rm spiral}<0.5$
    \item Spiral: $p_{\rm features\, or\, disk}>0.5$ and $p_{\rm spiral}>0.5$
\end{itemize}
where $p_{\rm [class]}>0.5$ indicates the debiased vote fraction for [class] from \citet{Hart+16}.  These thresholds were chosen to minimise the number of unclassifiable or ambiguous galaxies, and are therefore less stringent than those recommended by \citet{Willett+13}, and hence open to significant contamination between morphological classes.

\begin{figure}
    \centering
    \includegraphics[width=\columnwidth]{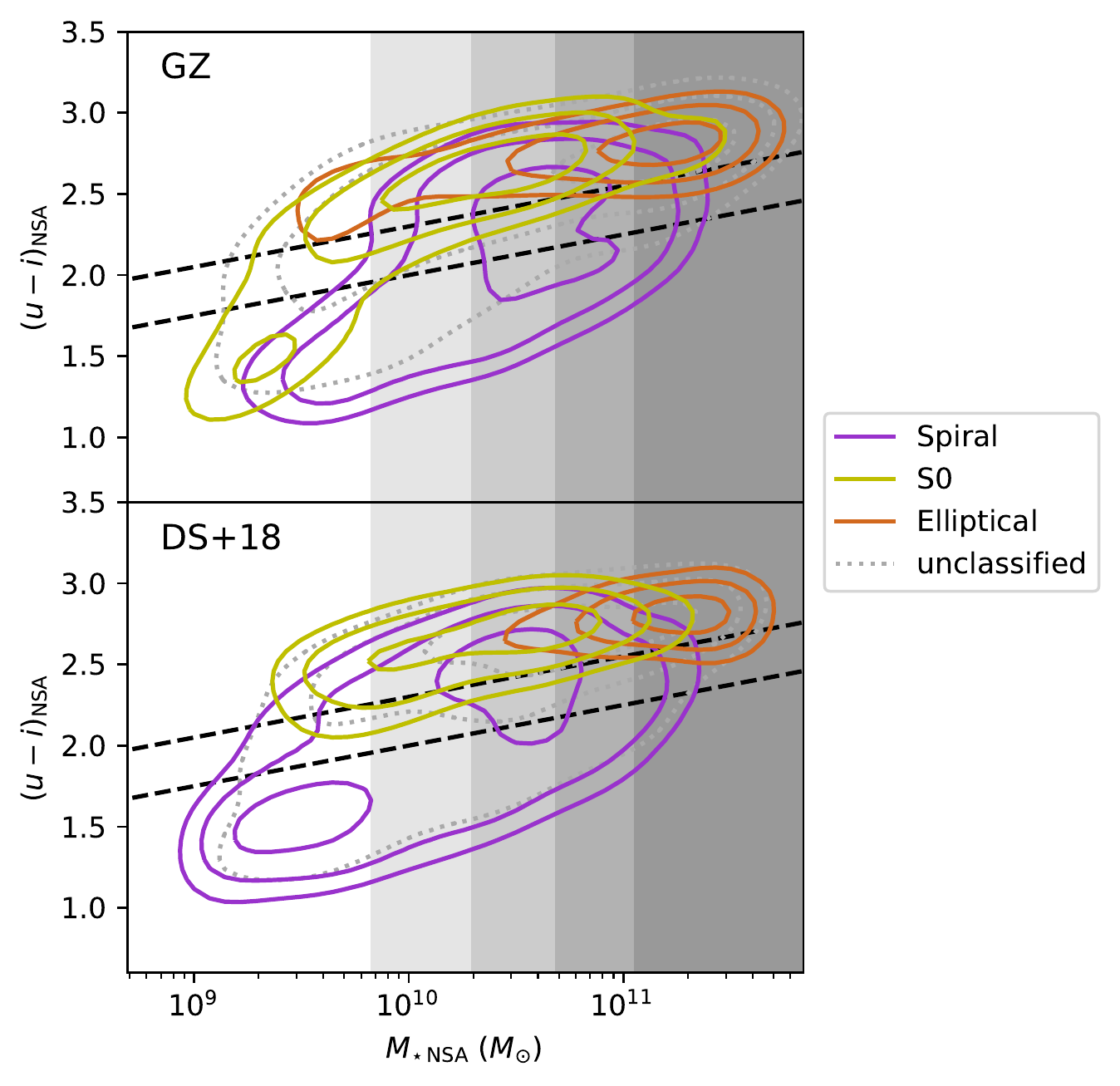}
    \caption{The distribution of galaxies in the stellar mass--colour plane (as Figure~\ref{fig:col-mass_col}), with galaxies coloured by their Galaxy Zoo (\textit{top}) and \citet{DominguezSanchez+18} machine learning (\textit{bottom}) morphological classifications as used in Section~\ref{sec:SFHoU-Morph}.  Spiral galaxies occupy the blue region of this parameter space in both schemes.  There is ambiguity at low stellar masses where the Galaxy Zoo classifications are less reliable due to difficulties in distinguishing spiral structure in SDSS imagery and the low thresholds used for separating morphologies.  Galaxies with low stellar mass are more likely to have spiral-morphology machine learning classifications compared to their Galaxy~Zoo classifications.  Contours of 30\%, 50\%, and 80\% of the peak unweighted density are shown for each classification.  For reference, dashed black lines delineate the boundaries between the colour classifications used in Section~\ref{sec:SFHoU-Col}, and shaded background regions indicate the stellar mass bins used throughout.}
    \label{fig:col-mass_gzml}
\end{figure}

The stellar mass--colour distribution of the Primary+ galaxies indicating their Galaxy~Zoo morphologies is shown in the upper panel of Figure~\ref{fig:col-mass_gzml}.  Over most of the sample's range in stellar mass, the spirals occupy the bluer region of parameter space, but at the low-mass ($<2\times 10^{9}\,~M_{\odot}$) end, blue galaxies are likely to have been classified as earlier-type galaxies, as we noted in \citet{Peterken+20Morph}.  This unexpected trend is likely because spiral structure in low-mass disk galaxies is harder to discern in SDSS imagery due to resolution effects.  The classifications we have implemented from the vote fractions to minimise the number of unclassified galaxies will therefore result in a high level of contamination between morphologies.  A lower spiral vote-fraction threshold might more reliably distinguish between morphologies in low-mass galaxies, but only at the expense of increasing the number of high-mass early-type disk galaxies being mis-classified as spirals.  Alternatively, using the recommended criteria of \citet{Willett+13} to obtain clean samples of galaxies of each morphology creates a large number of unclassified ``ambiguous'' galaxies, which when removed from the sample will render the MaNGA sample weightings described in Section~\ref{sec:Sample-weights} inaccurate.  However, we have repeated the analysis using debiased vote fraction thresholds of 80\% instead of 50\% in the criteria above and found no change to the results at high stellar masses, but at low stellar masses the dominance of galaxies with ``ambiguous'' morphologies makes any quantitative results impossible.


Alternative morphological classifications using the deep learning methods developed by \citet{DominguezSanchez+18} are available for galaxies which were part of the SDSS data release 15 (DR15; \citealt{DR15}).  This approach uses Galaxy Zoo and \citet{NairAbraham10} classifications as training sets to classify SDSS galaxy images.  Instead of replicating vote fractions for the presence of morphological features in each galaxy, the \citet{DominguezSanchez+18} approach directly provides a prediction for each galaxy's morphological T-type.  The resulting separation of different morphologies is cleaner at low stellar masses in the colour--mass plane, as illustrated in the lower panel of Figure~\ref{fig:col-mass_gzml}.  The question then arises of which classification method should be used.  Fortunately, in most cases we find that the results presented here do not change depending on which classification we use and we have therefore primarily presented results using the Galaxy~Zoo classifications to make use of the larger sample size.  In the instances where we find that the two classification methods provide conflicting results, we will describe and discuss those differences.

\subsubsection{Present-day morphology and the star-formation history of the Universe}
\label{sec:SFHoU-Morph-SFHoU}

\begin{figure}
    \centering
    \includegraphics[width=0.8\columnwidth]{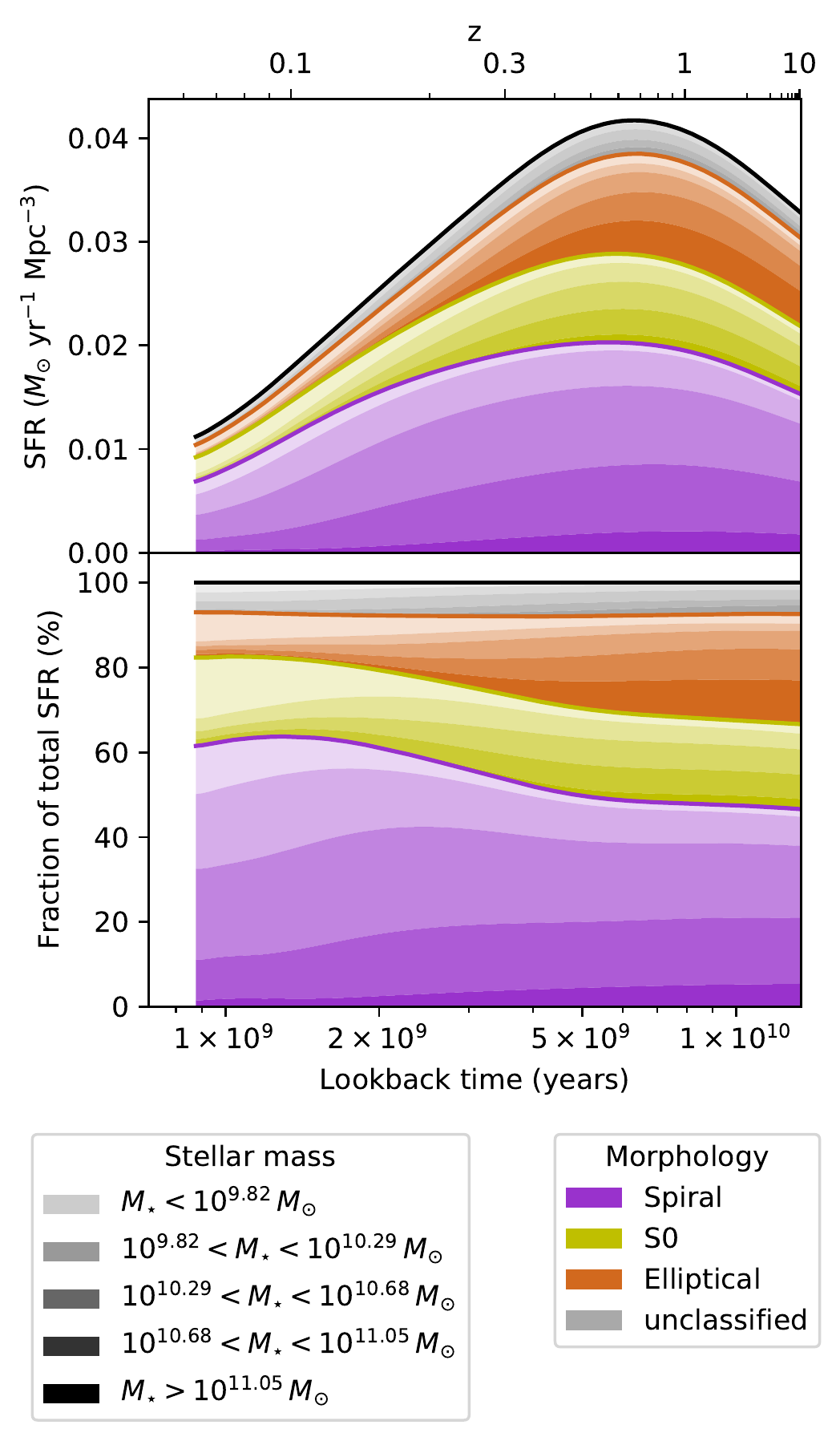}
    \caption{The star-formation history of the Universe inferred from the Primary+ sample, stratified into its contributions from galaxies in different stellar mass bins (shades; darkest to lightest for most- to least-massive galaxies) and of different Galaxy Zoo morphologies (colours), analogous to Figure~\ref{fig:Morphs_col}.  The \textit{top} panel shows the star-formation history, and the \textit{bottom} panel shows the percentage contribution from each of the sample's subdivisions to the total cosmic star-formation history.  Machine learning morphologies of \citet{DominguezSanchez+18} show similar results.}
    \label{fig:Morphs_gz}
\end{figure}

The cosmic star-formation history separated into contributions from galaxies of different present-day stellar masses and the Galaxy Zoo morphologies is shown in Figure~\ref{fig:Morphs_gz}.  Present-day spiral galaxies have contributed the greatest amount (at least 40\%) to the total star formation of any morphology at all lookback times, and contribute the majority (60\%) at the present day.  This 20\% morphology effect on a galaxy's past contribution to the star-formation history of the Universe is smaller than the 40\% colour increase seen in Figure~\ref{fig:Morphs_col}, implying that colour is a stronger indicator of a galaxy's full star-formation history than morphology, as found by others (e.g.\ \citealt{IbarraMedel+16}).

We find that a galaxy's morphology only reflects the last $2-3\,\textrm{Gyr}$ of its star-formation history on average, in that the relative contributions to the cosmic star-formation history by galaxies of different morphologies did not change until this lookback time.  We have previously demonstrated this result in more detail (see \citealt{Peterken+20Morph}).

\begin{figure*}
    \centering
    \includegraphics[width=\textwidth]{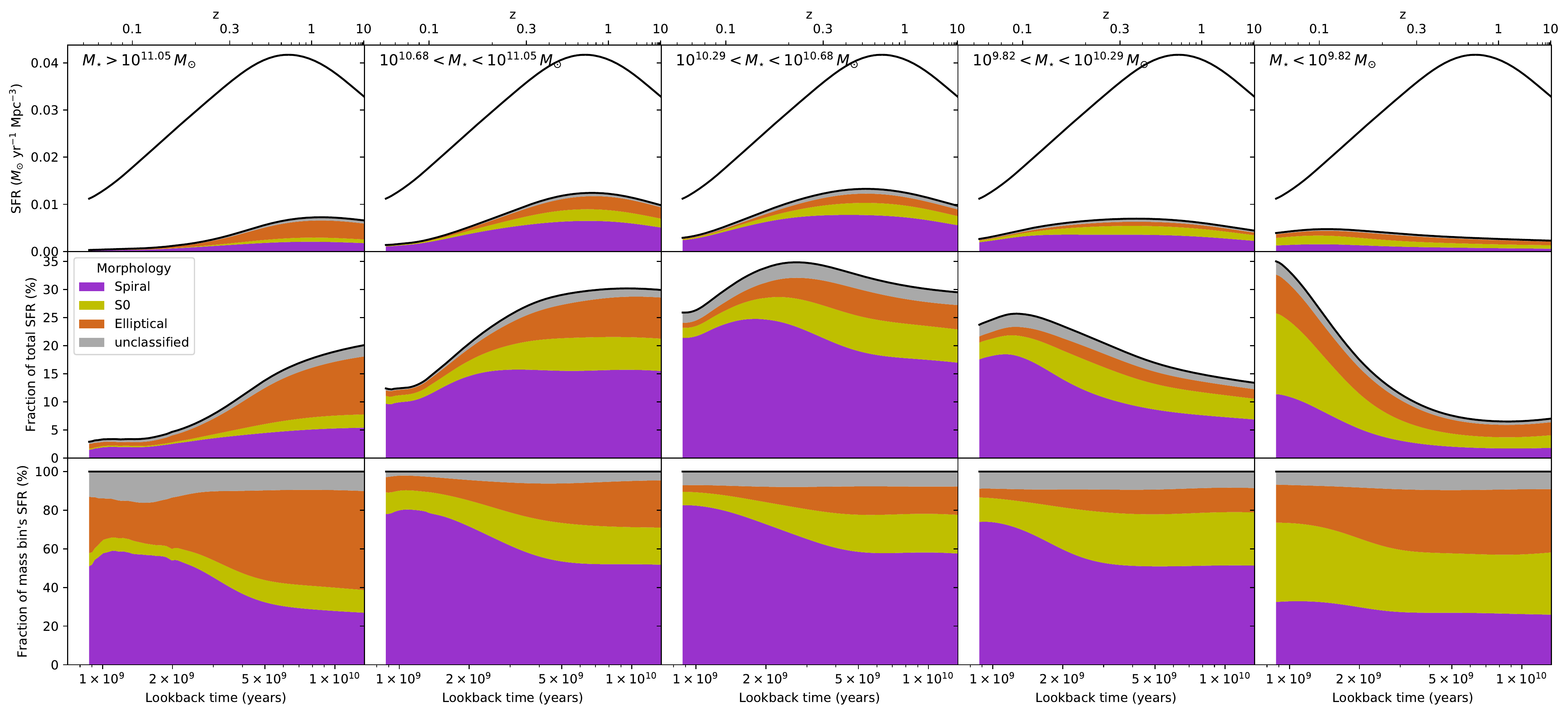}
    \caption{The absolute (\textit{top row}) and fractional (\textit{middle row}) contributions to the star-formation history of the Universe of galaxies of different morphologies in each bin of stellar mass (\textit{columns}), and the relative contribution of different Galaxy~Zoo morphologies to each mass bin's star-formation history (\textit{bottom row}).  Analogous to Figure~\ref{fig:MassSplits_col}.}
    \label{fig:MassSplits_gz}
\end{figure*}

However, morphology and mass are correlated, as shown in Figure~\ref{fig:col-mass_gzml}: observed morphology effects could be simply a result of early-type galaxies having higher stellar mass on average, and will therefore exhibit star-formation histories with higher star-formation rates at earlier lookback times than their late-type and less massive counterparts, as we saw in Figure~\ref{fig:PeaksBySample}.  To properly distinguish between downsizing and morphological effects, Figure~\ref{fig:MassSplits_gz} shows the same data as in Figure~\ref{fig:Morphs_gz} but with each mass bin separated.  We find that regardless of the morphological composition of the mass bin, spirals have increased their contribution to the star-formation history by $\sim20$\% between lookback times of 10 and 1~Gyr.  We therefore find the effects of present-day morphology on the contribution to the star-formation history of the Universe to be independent of present-day stellar mass.

However, we also see some evidence for the morphological classifications reflecting different timescales of star-formation histories in different mass bins in the same manner as we saw for the colour classifications in Figure~\ref{fig:MassSplits_col}.  Spiral galaxies with the largest present-day stellar mass began increasing in their contribution to their mass bin's star-formation history $\sim4\,\textrm{Gyr}$ ago to reach the current level $\sim2\,\textrm{Gyr}$ ago, compared to $\sim2\,\textrm{Gyr}$ ago for the lowest-mass spirals to start increasing their contribution with a trend of continued increase.

\begin{figure*}
    \centering
    \includegraphics[width=\textwidth]{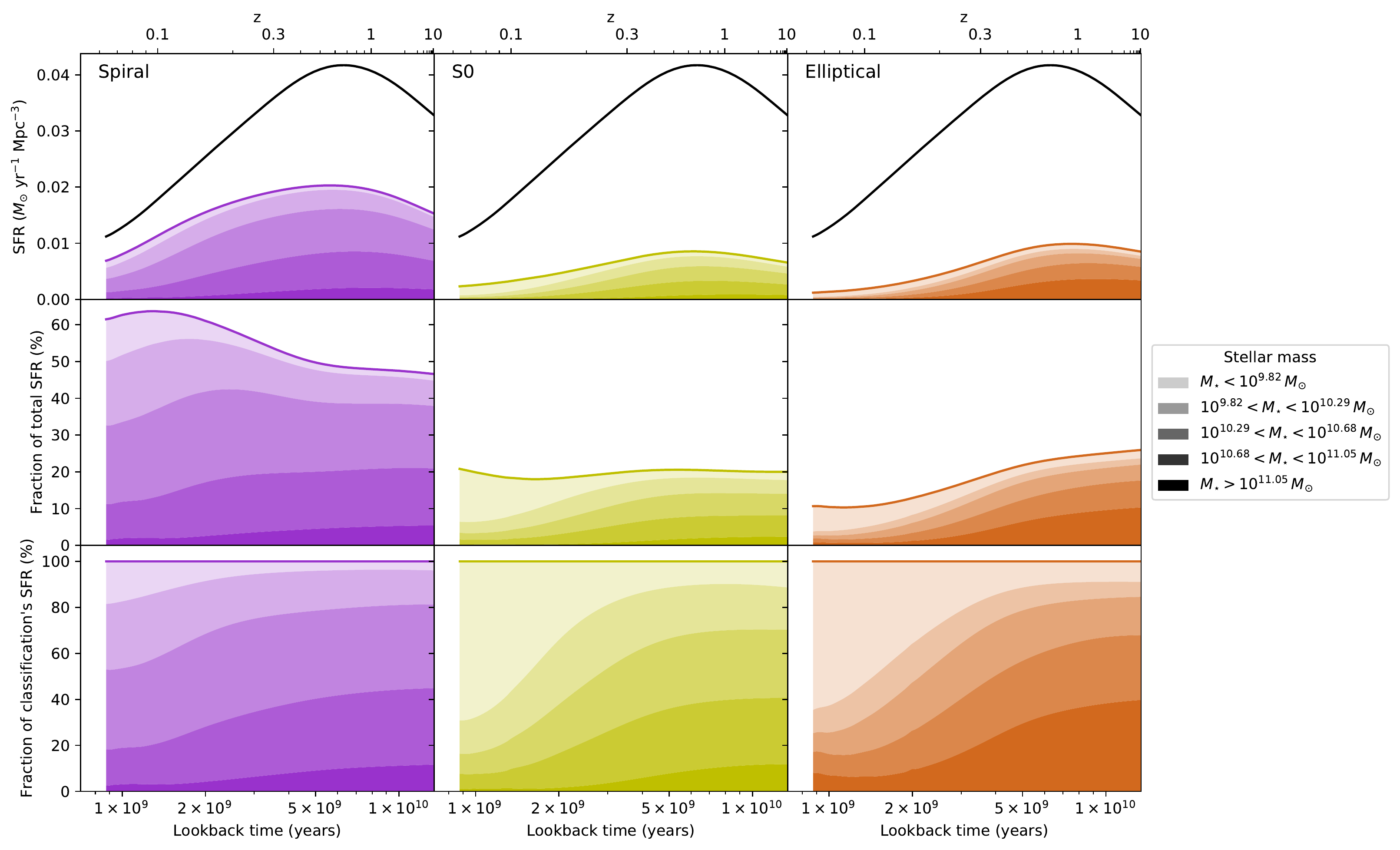}
    \caption{The absolute (\textit{top row}) and fractional (\textit{middle row}) contributions to the star-formation history of the Universe of galaxies with different present-day Galaxy Zoo moprhologies, stratified by the contributions from each stellar mass bin.  The relative contribution of galaxies with different stellar mass to each morphology's star-formation history is also shown (\textit{bottom~row}). Analogous to Figure~\ref{fig:MorphSplits_col}.}
    \label{fig:MorphSplits_gz}
\end{figure*}

As in Figure~\ref{fig:Morphs_col}, we are also able to reorganise the area plot of Figure~\ref{fig:Morphs_gz} by morphology, to show how galaxies of different present-day stellar masses have contributed to each morphological classification's star-formation history to reveal how the effects of stellar mass change among galaxies of different morphologies, as shown in Figure~\ref{fig:MorphSplits_gz}.  For the Galaxy Zoo morphologies, we find that downsizing effects are stronger in earlier-type galaxies; galaxies in the lowest-mass bin become more dominant among lenticular and elliptical galaxies at smaller lookback times than the equivalent effect in spiral galaxies.

\begin{figure*}
    \centering
    \includegraphics[width=\textwidth]{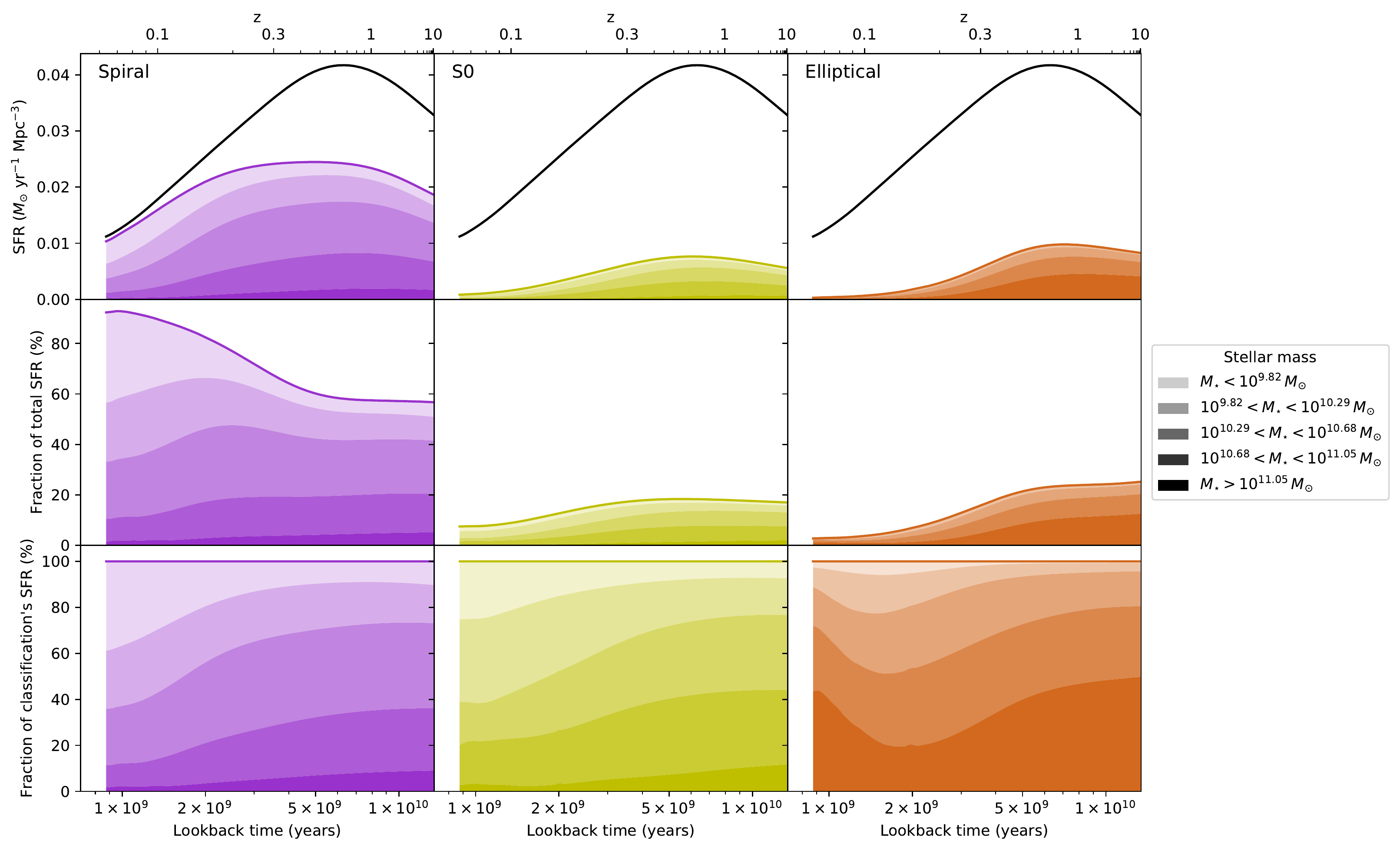}
    \caption{As Figure~\ref{fig:MorphSplits_gz} but using the \citet{DominguezSanchez+18} machine learning morphologies.  The galaxies which were not included in DR15 have been removed, and the weights of the remaining galaxies have been scaled such that the total weighted mass in the sample remains the same.  The total star-formation history for the \textit{full} Primary+ sample (black line) is shown, resulting in the sum of all shown contributions from each morphology to not equal exactly 100\% of the total at some lookback times.}
    \label{fig:MorphSplits_ml}
\end{figure*}

However, when using the \citet{DominguezSanchez+18} machine learning morphological classifications --- as shown in Figure~\ref{fig:MorphSplits_ml} --- we find the opposite; spiral galaxies show stronger downsizing effects than lenticular or elliptical galaxies.  We also find that elliptical galaxies have contributed almost no star formation in the last $\sim2\,\textrm{Gyr}$.  This difference highlights how carefully morphological classifications must be used to avoid biases associated with the difficulties inherent in identifying structures in low-mass galaxies.

The observed differences are easily explained by uncertainty in the classification of low-mass spiral galaxies, which appears to be more reliably resolved by the machine learning approach.  We wish to stress here that such a discrepancy merely highlights the importance of careful consideration of what exact morphological classifications are being used, and \textit{does not} imply that Galaxy~Zoo classifications are inherently flawed.  Instead, we argue that studies using any morphological classification method should be wary of how galaxies with potentially ambiguous morphologies are treated.  This is especially important if the classification scheme is based on a ``pure'' Hubble approach, which often contains significant subjectivity \citep{Naim+95} and assumptions \citep{Freeman70, Hart+17, Hart+18, Masters+19}, and does not necessarily reflect the physical processes occurring within galaxies \citep{Cappellari16, Wang+20}.  Indeed, we note again here that the implementation of morphological classifications using Galaxy Zoo vote fractions described in Section~\ref{sec:SFHoU-Morph-Classifications} goes contrary to the advice of \citet{Willett+13}, who detail how low-contamination samples of Galaxy Zoo galaxies with any required morphology can be effectively obtained.

In Figure~\ref{fig:MorphSplits_ml} (and to a lesser extent also in Figure~\ref{fig:MorphSplits_gz}), we see some apparent ``upsizing'' occurring in elliptical galaxies; high-mass galaxies appear to increase their relative contribution to the star-formation history of all ellipticals in the last $\sim2\,\textrm{Gyr}$.  This phenomenon is likely due to the problems that we addressed in \citet{Peterken+20FR} of the ``UV~upturn'' (see also \citealt{Yi08, CidFernandes+GonzalezDelgado10}) being most dominant in high-mass ellipticals due to their extremely old stellar populations.  That this effect is strongest using the \citet{DominguezSanchez+18} morphological classifications is therefore due to the population of possibly misclassified low-mass galaxies in the Galaxy~Zoo classifications diluting the effect with ``real'' star formation; see also \citet[Section~4.4]{Fischer+19} who describe this phenomenon in more detail.  However, we note that the observed effects are extremely small; even with this spurious increase of star-formation rate in high-mass elliptical galaxies, those galaxies still only account for $<1\%$ of the star formation in the present-day Universe regardless of which morphological classification method is used, so these effects do not significantly affect the derived cosmic star-formation history.

We find the total star-formation histories of all galaxies with stellar mass $M_{\star}>10^{10.29}\,M_{\odot}$ (i.e.\ in the three bins of greatest stellar mass) to be similar for lenticular and elliptical classifications.  This similarity could imply that all massive present-day early-type galaxies have experienced similar star-formation histories, with the difference in morphologies due to other factors such as merger rate histories.  Alternatively, it could be evidence for visually-determined shape being an ineffective way to separate the bimodality of fast- and slow-rotating early-type galaxies \citep{Emsellem+07, Cappellari+11, Cappellari16, Graham+18}, or a combination of both of these effects.

\begin{figure}
    \centering
    \includegraphics[width=0.9\columnwidth]{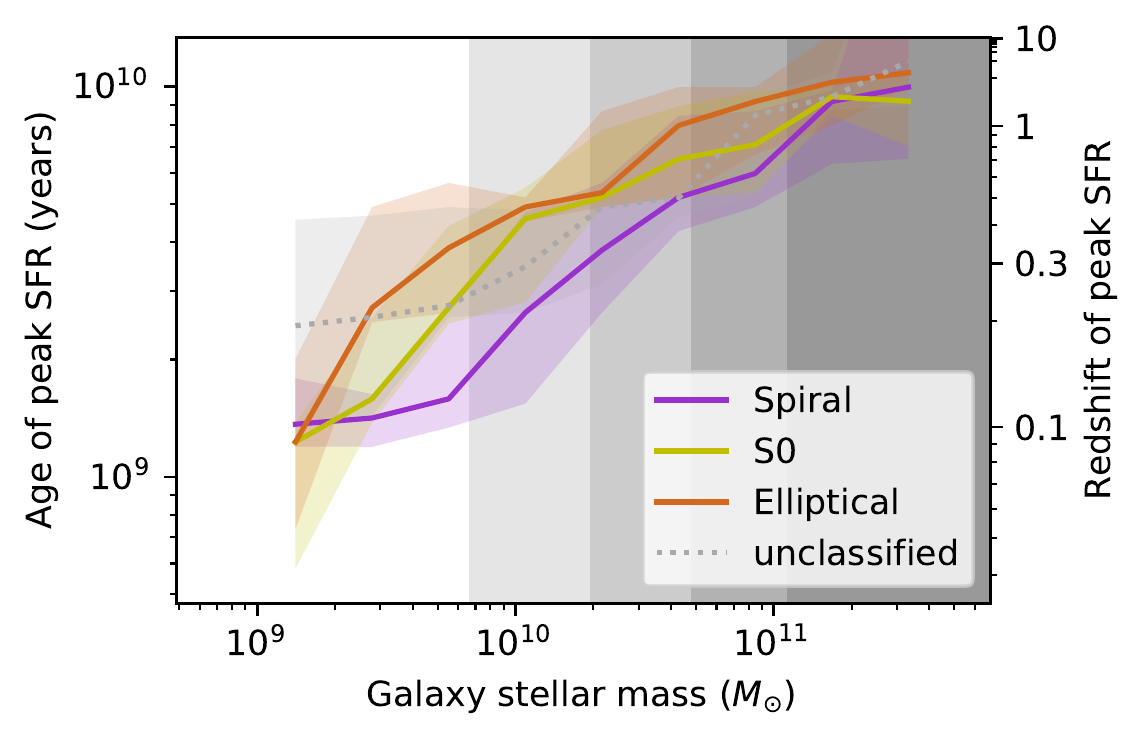}
    \caption{The peak in star-formation history of galaxies of each present-day morphological classification in the Primary+ sample, as determined using Galaxy~Zoo.  Solid lines indicate the weighted medians and the coloured shaded regions enclose the area between the one- and two-thirds weighted percentiles.}
    \label{fig:PeaksByMorph_gz}
\end{figure}

\begin{figure}
    \centering
    \includegraphics[width=0.9\columnwidth]{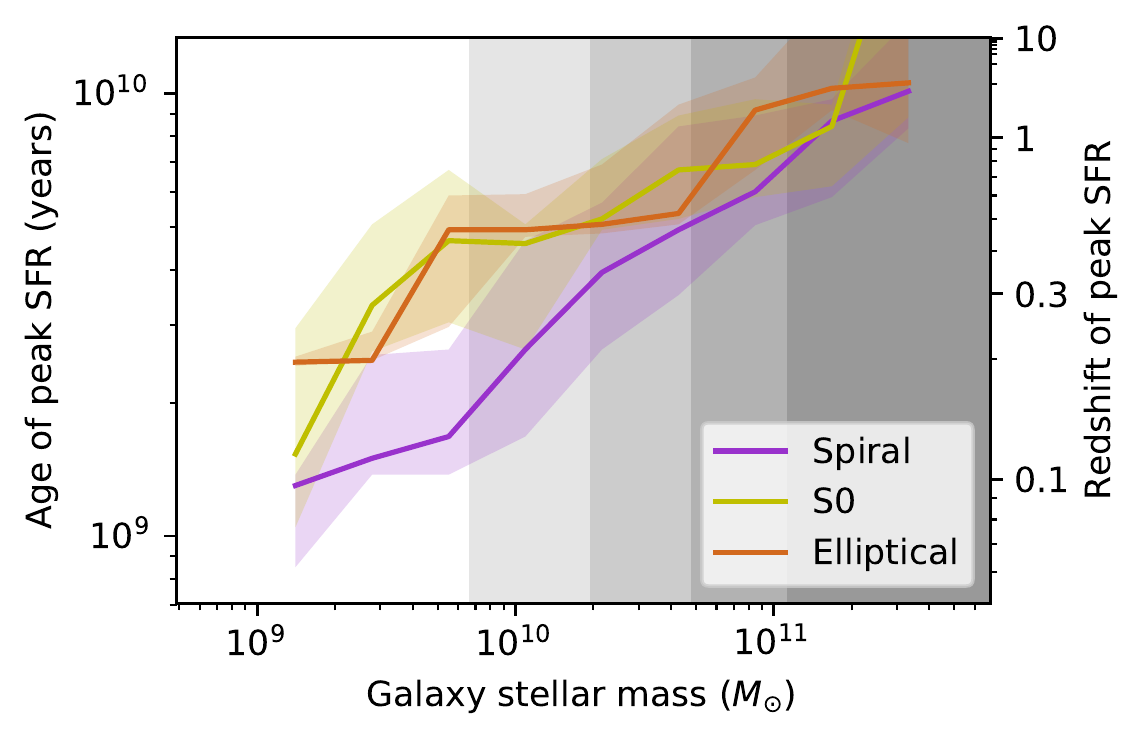}
    \caption{As Figure~\ref{fig:PeaksByMorph_gz}, but using the \citet{DominguezSanchez+18} machine learning morphological classifications of DR15 galaxies.}
    \label{fig:PeaksByMorph_ml}
\end{figure}

\begin{figure}
    \centering
    \includegraphics[width=0.9\columnwidth]{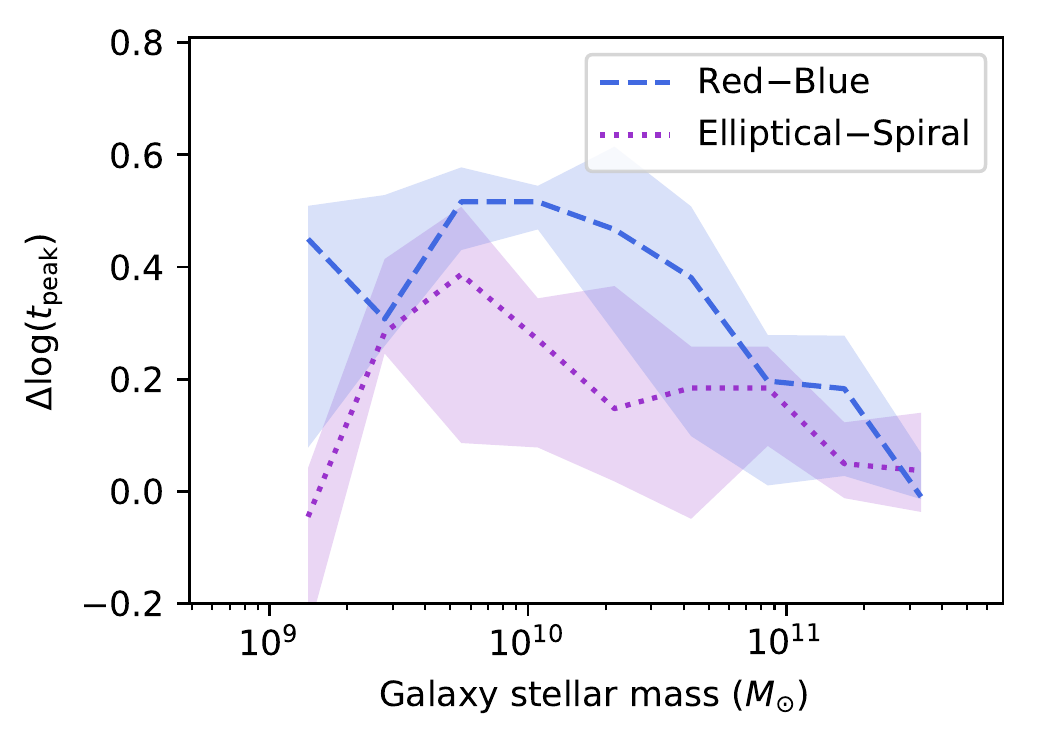}
    \caption{The offset $\Delta \log\left(t_{\rm peak}\right)$ in lookback time of peak star-formation rate $t_{\rm peak}$ observed between present-day red sequence and blue cloud galaxies (blue dashed line), and between present-day elliptical and spiral galaxies (purple dotted line) using Galaxy~Zoo classifications, as a function of present-day galaxy stellar mass $M_{\star}$.  The effect of present-day colour is larger than that of present-day morphology at all stellar masses, but is particularly significant in the range $7 \times 10^{9} M_{\odot} \lessapprox M_\star \lessapprox 4 \times 10^{10} M_{\odot}$.}
    \label{fig:PeaksDifference}
\end{figure}

By comparing the lookback time of peak star formation of galaxies with different present-day morphologies in Figures~\ref{fig:PeaksByMorph_gz} (for Galaxy Zoo morphologies) and \ref{fig:PeaksByMorph_ml} (for \citet{DominguezSanchez+18} machine learning morphologies), we see that spiral galaxies peaked more recently than early-type galaxies at all present-day stellar masses regardless of the classification method.  We find that this difference is greater in galaxies with low stellar mass, and is greatest around $M_{\star}=6\times10^{9}\,M_{\odot}$; above this threshold we find that both classification schemes display stronger downsizing among spirals than in early-type galaxies, in agreement with \citet{Goddard+17}.

We also find that the peak in star-formation history of lenticular and elliptical galaxies is similar for early-type galaxies with stellar masses $M_{\star}>\sim10^{10}\,M_{\odot}$ using the Galaxy Zoo classifications (Figure~\ref{fig:PeaksByMorph_gz}) and  for stellar masses $M_{\star}>\sim10^{9}\,M_{\odot}$ using the \citet{DominguezSanchez+18} classifications (Figure~\ref{fig:PeaksByMorph_ml}), again reflecting either the similar histories of these galaxies or the difficulty in distinguishing between them through imagery alone.  We see that the star-formation peak in present-day lenticular galaxies classified by Galaxy Zoo becomes systematically more recent for lower stellar masses, highlighting likely contamination of mis-classified spiral galaxies using the thresholds applied here.

However, regardless of which classification method is used, we see again that the effect of morphology on a galaxy's contribution to the cosmic star-formation history is smaller than the effect of colour: the difference in peak star-formation times between early- and late-type galaxies is smaller than the difference between blue cloud and red sequence galaxies ($\sim0.2$~dex and $\sim0.5$~dex respectively at $M_{\star} = 10^{10}\,M_{\odot}$ for example).  This is illustrated at all present-day stellar masses in Figure~\ref{fig:PeaksDifference}.

\section{Size versus shade versus shape: Conclusions and interpretation}
\label{sec:Interpretation}

Here, we have recovered evidence for downsizing, in that the fraction of star-formation occurring in the galaxies with greatest present-day stellar mass was largest at early times, with low-mass galaxies dominating the cosmic star-formation at more recent times, and that the lookback time corresponding to a galaxy's peak star-formation rate is therefore correlated with its present-day stellar mass.  We find that this correlation exhibits the strongest gradient among low-mass galaxies, suggesting accelerating downsizing at more recent times.  By further splitting the galaxy sample, we subsequently showed that downsizing effects are present in galaxies of all present-day colours (which we showed to be analogous to specific star-formation rate) and morphologies.  A galaxy's present-day stellar mass is therefore a significant indicator of its star-formation history regardless of its other properties.

In quantifying these effects, we found that galaxies currently in the blue cloud exhibited stronger downsizing effects than those in the green valley or red sequence.  However, we found some differences in whether downsizing is strongest in present-day early- or late-type galaxies depending on which classification method was used, which we argue is due to our non-standard implementation of Galaxy Zoo morphologies, and therefore highlights the care that must be taken when separating galaxies by their visual morphologies.  Reassuringly, there is agreement that late-type galaxies exhibit the stronger downsizing effects among galaxies with stellar mass $M_{\star}>6\times10^{9}\,M_{\odot}$ and that high-mass elliptical and lenticular galaxies have had similar star-formation histories.

Irrespective of classification scheme, we find that the effect of present-day colour is greater than the effect of present-day morphology on a galaxy's star-formation history --- confirming previous results \citep{IbarraMedel+16, GarciaBenito+19} --- in that the contribution from present-day blue cloud galaxies increased by $\sim40\%$ over the past $10\,\textrm{Gyr}$ but that from spirals only increased by $\sim20\%$ over the same time period by comparison.  Similarly, although we found that present-day spirals and blue cloud galaxies exhibited the more recent peaks in star-formation at all stellar masses, the typical difference between spiral and elliptical galaxies ($\sim0.3\,\textrm{dex}$) is less than that between blue cloud and red sequence galaxies ($\sim0.6\,\textrm{dex}$).  These results suggest that galaxies of similar morphologies are more likely to have undergone more significantly different histories compared to the equivalent variation in histories among galaxies of similar colours.

However, we also find evidence that these straightforward results do not tell the complete story.  Specifically, we find that both present-day colour and morphology reflect only the more recent star-formation histories among lower-mass galaxies compared to their counterparts with high stellar mass.  We interpret this as reflective of a galaxy's inertia to change.  The physical processes required to suppress or rejuvenate star-formation or to restructure a galaxy must occur over longer timescales to significantly alter a massive galaxy's observed properties.  Meanwhile, the tidal, stripping, starvation, or inflow processes experienced by less massive galaxies are able to affect their star-formation rates (and therefore colours) and morphologies much more rapidly.

We have therefore demonstrated that a galaxy's mass, colour and morphology --- size, shade and shape --- all indicate its historical contribution to the cosmic star-formation history, but to different extents and with codependencies such that all three properties must be considered to build a full picture.

\section{Summary}

Using an established stellar population ``fossil record'' analysis, we obtained star-formation histories of the inner $1.2\,R_{\rm e}$ of 6861 galaxies from the SDSS-IV MaNGA survey, with 2519 galaxies also sampled to $2.3\,R_{\rm e}$.  By carefully weighting each galaxy to create an effectively volume-limited sample, we inferred the star-formation history of the Universe, which was found to be in general agreement with previous observations of galaxy populations at different redshifts.

We showed that a galaxy's star-formation history is linked to its present-day mass, its colour, and its morphology, although colour is more strongly connected than morphology regarding historical contributions to the cosmic star-formation history.  We also found evidence for downsizing effects being significant in galaxies of all present-day colours and morphologies, in that low-mass galaxies of all types have become more dominant in their contribution to the cosmic star-formation rate in more recent times.  These effects, however, were found to be most significant among galaxies currently in the blue cloud.  Different morphological classification schemes gave contradictory results on whether downsizing is strongest among galaxies designated as being currently early or late type, but show agreement that spiral galaxies exhibit the stronger effects among galaxies with present-day stellar mass $M_{\star}>6\times10^{9}\,M_{\odot}$.

We also found that the historical contribution to the cosmic star-formation history from galaxies currently in the ``green valley'' does not directly follow that of either the blue cloud or the red sequence populations, with these galaxies' star-formation histories being remarkably representative of that of the Universe as a whole.

These results once again demonstrate the power of stellar population fossil record techniques in uncovering the link between the Universe's past with the galaxies we see today.

\section{Data Availability}

This publication uses the team-internal MPL-9 version of the MaNGA science data products, which are broadly similar to previous versions available publicly through SDSS Data Releases DR13 \citep{DR13}, DR14 \citep{DR14}, and DR15 \citep{DR15}.  Comparable data products containing the full MaNGA galaxy sample --- including the full sample used here --- will be publicly released in 2021 as part of SDSS DR17, as will the raw data and all previous versions of the data reduction pipeline.

\section*{Acknowledgements}

The authors thank the anonymous reviewer for their helpful suggestions and improvements to this paper.

The authors also thank S.\ F.\ S{\'a}nchez, D.\ Wake, A.\ R.\ Calette and A.\ Rodriguez-Puebla for their extensive help and support on the technical aspects of this work.

Funding for the Sloan Digital Sky Survey IV has been provided by the Alfred P. Sloan Foundation, the U.S. Department of Energy Office of Science, and the Participating Institutions. SDSS acknowledges support and resources from the Center for High-Performance Computing at the University of Utah. The SDSS web site is \url{www.sdss.org}.

SDSS is managed by the Astrophysical Research Consortium for the Participating Institutions of the SDSS Collaboration including the Brazilian Participation Group, the Carnegie Institution for Science, Carnegie Mellon University, the Chilean Participation Group, the French Participation Group, Harvard-Smithsonian Center for Astrophysics, Instituto de Astrof{\'i}sica de Canarias, The Johns Hopkins University, Kavli Institute for the Physics and Mathematics of the Universe (IPMU) / University of Tokyo, the Korean Participation Group, Lawrence Berkeley National Laboratory, Leibniz Institut f{\:u}r Astrophysik Potsdam (AIP), Max-Planck-Institut f{\:u}r Astronomie (MPIA Heidelberg), Max-Planck-Institut f{\:u}r Astrophysik (MPA Garching), Max-Planck-Institut f{\:u}r Extraterrestrische Physik (MPE), National Astronomical Observatories of China, New Mexico State University, New York University, University of Notre Dame, Observat{\'o}rio Nacional / MCTI, The Ohio State University, Pennsylvania State University, Shanghai Astronomical Observatory, United Kingdom Participation Group, Universidad Nacional Aut{\'o}noma de M{\'e}xico, University of Arizona, University of Colorado Boulder, University of Oxford, University of Portsmouth, University of Utah, University of Virginia, University of Washington, University of Wisconsin, Vanderbilt University, and Yale University.

This publication uses data generated via the Zooniverse.org platform, development of which is funded by generous support, including a Global Impact Award from Google, and by a grant from the Alfred P. Sloan Foundation.

We are grateful for access to the University of Nottingham's \texttt{Augusta} high performance computing facility.




\bibliographystyle{mnras}
\bibliography{refs} 

\begin{thebibliography}{}
\makeatletter
\relax
\def\mn@urlcharsother{\let\do\@makeother \do\$\do\&\do\#\do\^\do\_\do\%\do\~}
\def\mn@doi{\begingroup\mn@urlcharsother \@ifnextchar [ {\mn@doi@}
  {\mn@doi@[]}}
\def\mn@doi@[#1]#2{\def\@tempa{#1}\ifx\@tempa\@empty \href
  {http://dx.doi.org/#2} {doi:#2}\else \href {http://dx.doi.org/#2} {#1}\fi
  \endgroup}
\def\mn@eprint#1#2{\mn@eprint@#1:#2::\@nil}
\def\mn@eprint@arXiv#1{\href {http://arxiv.org/abs/#1} {{\tt arXiv:#1}}}
\def\mn@eprint@dblp#1{\href {http://dblp.uni-trier.de/rec/bibtex/#1.xml}
  {dblp:#1}}
\def\mn@eprint@#1:#2:#3:#4\@nil{\def\@tempa {#1}\def\@tempb {#2}\def\@tempc
  {#3}\ifx \@tempc \@empty \let \@tempc \@tempb \let \@tempb \@tempa \fi \ifx
  \@tempb \@empty \def\@tempb {arXiv}\fi \@ifundefined
  {mn@eprint@\@tempb}{\@tempb:\@tempc}{\expandafter \expandafter \csname
  mn@eprint@\@tempb\endcsname \expandafter{\@tempc}}}

\bibitem[\protect\citeauthoryear{{Abolfathi} et~al.,}{{Abolfathi}
  et~al.}{2018}]{DR14}
{Abolfathi} B.,  et~al., 2018, \mn@doi [\apjs] {10.3847/1538-4365/aa9e8a},
  \href {https://ui.adsabs.harvard.edu/abs/2018ApJS..235...42A} {235, 42}

\bibitem[\protect\citeauthoryear{{Aguado} et~al.,}{{Aguado}
  et~al.}{2019}]{DR15}
{Aguado} D.~S.,  et~al., 2019, \mn@doi [\apjs] {10.3847/1538-4365/aaf651},
  \href {https://ui.adsabs.harvard.edu/abs/2019ApJS..240...23A} {240, 23}

\bibitem[\protect\citeauthoryear{{Albareti} et~al.,}{{Albareti}
  et~al.}{2017}]{DR13}
{Albareti} F.~D.,  et~al., 2017, \mn@doi [\apjs] {10.3847/1538-4365/aa8992},
  \href {https://ui.adsabs.harvard.edu/abs/2017ApJS..233...25A} {233, 25}

\bibitem[\protect\citeauthoryear{{Asa'd}, {Vazdekis}, {Cervi{\~n}o},
  {No{\"e}l}, {Beasley}  \& {Kassab}}{{Asa'd} et~al.}{2017}]{Asa'd+17}
{Asa'd} R.~S.,  {Vazdekis} A.,  {Cervi{\~n}o} M.,  {No{\"e}l} N. E.~D.,
  {Beasley} M.~A.,   {Kassab} M.,  2017, \mn@doi [\mnras]
  {10.1093/mnras/stx1824}, \href
  {https://ui.adsabs.harvard.edu/abs/2017MNRAS.471.3599A} {471, 3599}

\bibitem[\protect\citeauthoryear{{Behroozi}, {Wechsler}  \&
  {Conroy}}{{Behroozi} et~al.}{2013}]{Behroozi+13}
{Behroozi} P.~S.,  {Wechsler} R.~H.,   {Conroy} C.,  2013, \mn@doi [\apj]
  {10.1088/0004-637X/770/1/57}, \href
  {https://ui.adsabs.harvard.edu/abs/2013ApJ...770...57B} {770, 57}

\bibitem[\protect\citeauthoryear{{Belfiore} et~al.,}{{Belfiore}
  et~al.}{2018}]{Belfiore+18}
{Belfiore} F.,  et~al., 2018, \mn@doi [\mnras] {10.1093/mnras/sty768}, \href
  {https://ui.adsabs.harvard.edu/abs/2018MNRAS.477.3014B} {477, 3014}

\bibitem[\protect\citeauthoryear{{Belfiore} et~al.,}{{Belfiore}
  et~al.}{2019}]{DAPLines}
{Belfiore} F.,  et~al., 2019, \mn@doi [\aj] {10.3847/1538-3881/ab3e4e}, \href
  {https://ui.adsabs.harvard.edu/abs/2019AJ....158..160B} {158, 160}

\bibitem[\protect\citeauthoryear{{Bell} et~al.,}{{Bell} et~al.}{2012}]{Bell+12}
{Bell} E.~F.,  et~al., 2012, \mn@doi [\apj] {10.1088/0004-637X/753/2/167},
  \href {https://ui.adsabs.harvard.edu/abs/2012ApJ...753..167B} {753, 167}

\bibitem[\protect\citeauthoryear{{Bellstedt} et~al.,}{{Bellstedt}
  et~al.}{2020}]{Bellstedt+20}
{Bellstedt} S.,  et~al., 2020, arXiv e-prints, \href
  {https://ui.adsabs.harvard.edu/abs/2020arXiv200511917B} {p. arXiv:2005.11917}

\bibitem[\protect\citeauthoryear{{Blanton} \& {Moustakas}}{{Blanton} \&
  {Moustakas}}{2009}]{BlantonMoustakas09}
{Blanton} M.~R.,  {Moustakas} J.,  2009, \mn@doi [\araa]
  {10.1146/annurev-astro-082708-101734}, \href
  {https://ui.adsabs.harvard.edu/abs/2009ARA&A..47..159B} {47, 159}

\bibitem[\protect\citeauthoryear{{Blanton}, {Kazin}, {Muna}, {Weaver}  \&
  {Price-Whelan}}{{Blanton} et~al.}{2011}]{NSA}
{Blanton} M.~R.,  {Kazin} E.,  {Muna} D.,  {Weaver} B.~A.,   {Price-Whelan} A.,
   2011, \mn@doi [AJ] {10.1088/0004-6256/142/1/31}, \href
  {http://adsabs.harvard.edu/abs/2011AJ....142...31B} {142, 31}

\bibitem[\protect\citeauthoryear{{Blanton} et~al.,}{{Blanton}
  et~al.}{2017}]{Blanton+17}
{Blanton} M.~R.,  et~al., 2017, \mn@doi [AJ] {10.3847/1538-3881/aa7567}, \href
  {http://adsabs.harvard.edu/abs/2017AJ....154...28B} {154, 28}

\bibitem[\protect\citeauthoryear{{Bluck}, {Mendel}, {Ellison}, {Moreno},
  {Simard}, {Patton}  \& {Starkenburg}}{{Bluck} et~al.}{2014}]{Bluck+14}
{Bluck} A. F.~L.,  {Mendel} J.~T.,  {Ellison} S.~L.,  {Moreno} J.,  {Simard}
  L.,  {Patton} D.~R.,   {Starkenburg} E.,  2014, \mn@doi [\mnras]
  {10.1093/mnras/stu594}, \href
  {https://ui.adsabs.harvard.edu/abs/2014MNRAS.441..599B} {441, 599}

\bibitem[\protect\citeauthoryear{{Bremer} et~al.,}{{Bremer}
  et~al.}{2018}]{Bremer+18}
{Bremer} M.~N.,  et~al., 2018, \mn@doi [\mnras] {10.1093/mnras/sty124}, \href
  {https://ui.adsabs.harvard.edu/abs/2018MNRAS.476...12B} {476, 12}

\bibitem[\protect\citeauthoryear{{Bundy} et~al.,}{{Bundy}
  et~al.}{2015}]{Bundy+15}
{Bundy} K.,  et~al., 2015, \mn@doi [ApJ] {10.1088/0004-637X/798/1/7}, \href
  {http://adsabs.harvard.edu/abs/2015ApJ...798....7B} {798, 7}

\bibitem[\protect\citeauthoryear{{Calzetti}, {Armus}, {Bohlin}, {Kinney},
  {Koornneef}  \& {Storchi-Bergmann}}{{Calzetti} et~al.}{2000}]{Calzetti+00}
{Calzetti} D.,  {Armus} L.,  {Bohlin} R.~C.,  {Kinney} A.~L.,  {Koornneef} J.,
   {Storchi-Bergmann} T.,  2000, \mn@doi [\apj] {10.1086/308692}, \href
  {https://ui.adsabs.harvard.edu/abs/2000ApJ...533..682C} {533, 682}

\bibitem[\protect\citeauthoryear{{Cappellari}}{{Cappellari}}{2016}]{Cappellari16}
{Cappellari} M.,  2016, \mn@doi [\araa] {10.1146/annurev-astro-082214-122432},
  \href {https://ui.adsabs.harvard.edu/abs/2016ARA&A..54..597C} {54, 597}

\bibitem[\protect\citeauthoryear{{Cappellari} et~al.,}{{Cappellari}
  et~al.}{2011}]{Cappellari+11}
{Cappellari} M.,  et~al., 2011, \mn@doi [\mnras]
  {10.1111/j.1365-2966.2011.18600.x}, \href
  {https://ui.adsabs.harvard.edu/abs/2011MNRAS.416.1680C} {416, 1680}

\bibitem[\protect\citeauthoryear{{Catal{\'a}n-Torrecilla}
  et~al.,}{{Catal{\'a}n-Torrecilla} et~al.}{2015}]{CatalanTorrecilla+15}
{Catal{\'a}n-Torrecilla} C.,  et~al., 2015, \mn@doi [\aap]
  {10.1051/0004-6361/201526023}, \href
  {https://ui.adsabs.harvard.edu/abs/2015A&A...584A..87C} {584, A87}

\bibitem[\protect\citeauthoryear{{Chabrier}}{{Chabrier}}{2003}]{Chabrier03}
{Chabrier} G.,  2003, \mn@doi [\pasp] {10.1086/376392}, \href
  {https://ui.adsabs.harvard.edu/abs/2003PASP..115..763C} {115, 763}

\bibitem[\protect\citeauthoryear{{Charlot}, {Worthey}  \& {Bressan}}{{Charlot}
  et~al.}{1996}]{Charlot+96}
{Charlot} S.,  {Worthey} G.,   {Bressan} A.,  1996, \mn@doi [\apj]
  {10.1086/176759}, \href
  {https://ui.adsabs.harvard.edu/abs/1996ApJ...457..625C} {457, 625}

\bibitem[\protect\citeauthoryear{{Cherinka} et~al.,}{{Cherinka}
  et~al.}{2019}]{Marvin}
{Cherinka} B.,  et~al., 2019, \mn@doi [\aj] {10.3847/1538-3881/ab2634}, \href
  {https://ui.adsabs.harvard.edu/abs/2019AJ....158...74C} {158, 74}

\bibitem[\protect\citeauthoryear{{Cheung} et~al.,}{{Cheung}
  et~al.}{2012}]{Cheung+12}
{Cheung} E.,  et~al., 2012, \mn@doi [\apj] {10.1088/0004-637X/760/2/131}, \href
  {https://ui.adsabs.harvard.edu/abs/2012ApJ...760..131C} {760, 131}

\bibitem[\protect\citeauthoryear{{Cid Fernandes}}{{Cid
  Fernandes}}{2018}]{FakeNews}
{Cid Fernandes} R.,  2018, \mn@doi [\mnras] {10.1093/mnras/sty2012}, \href
  {http://adsabs.harvard.edu/abs/2018MNRAS.480.4480C} {480, 4480}

\bibitem[\protect\citeauthoryear{{Cid Fernandes} \& {Gonz{\'a}lez
  Delgado}}{{Cid Fernandes} \& {Gonz{\'a}lez
  Delgado}}{2010}]{CidFernandes+GonzalezDelgado10}
{Cid Fernandes} R.,  {Gonz{\'a}lez Delgado} R.~M.,  2010, \mn@doi [\mnras]
  {10.1111/j.1365-2966.2009.16153.x}, \href
  {https://ui.adsabs.harvard.edu/abs/2010MNRAS.403..780C} {403, 780}

\bibitem[\protect\citeauthoryear{{Cid Fernandes}, {Mateus}, {Sodr{\'e}},
  {Stasi{\'n}ska}  \& {Gomes}}{{Cid Fernandes} et~al.}{2005}]{Starlight}
{Cid Fernandes} R.,  {Mateus} A.,  {Sodr{\'e}} L.,  {Stasi{\'n}ska} G.,
  {Gomes} J.~M.,  2005, \mn@doi [MNRAS] {10.1111/j.1365-2966.2005.08752.x},
  \href {http://adsabs.harvard.edu/abs/2005MNRAS.358..363C} {358, 363}

\bibitem[\protect\citeauthoryear{{Cid Fernandes} et~al.,}{{Cid Fernandes}
  et~al.}{2013}]{CidFernandes+13}
{Cid Fernandes} R.,  et~al., 2013, \mn@doi [\aap]
  {10.1051/0004-6361/201220616}, \href
  {http://adsabs.harvard.edu/abs/2013A%26A...557A..86C} {557, A86}

\bibitem[\protect\citeauthoryear{{Cid Fernandes} et~al.,}{{Cid Fernandes}
  et~al.}{2014}]{CidFernandes+14}
{Cid Fernandes} R.,  et~al., 2014, \mn@doi [\aap]
  {10.1051/0004-6361/201321692}, \href
  {http://adsabs.harvard.edu/abs/2014A%26A...561A.130C} {561, A130}

\bibitem[\protect\citeauthoryear{{Connolly}, {Szalay}, {Dickinson}, {SubbaRao}
  \& {Brunner}}{{Connolly} et~al.}{1997}]{Connolly+97}
{Connolly} A.~J.,  {Szalay} A.~S.,  {Dickinson} M.,  {SubbaRao} M.~U.,
  {Brunner} R.~J.,  1997, \mn@doi [\apjl] {10.1086/310829}, \href
  {https://ui.adsabs.harvard.edu/abs/1997ApJ...486L..11C} {486, L11}

\bibitem[\protect\citeauthoryear{{Conroy} \& {Gunn}}{{Conroy} \&
  {Gunn}}{2010}]{ConroyGunn10}
{Conroy} C.,  {Gunn} J.~E.,  2010, \mn@doi [\apj]
  {10.1088/0004-637X/712/2/833}, \href
  {https://ui.adsabs.harvard.edu/abs/2010ApJ...712..833C} {712, 833}

\bibitem[\protect\citeauthoryear{{Conroy}, {Gunn}  \& {White}}{{Conroy}
  et~al.}{2009}]{Conroy+09}
{Conroy} C.,  {Gunn} J.~E.,   {White} M.,  2009, \mn@doi [ApJ]
  {10.1088/0004-637X/699/1/486}, \href
  {http://adsabs.harvard.edu/abs/2009ApJ...699..486C} {699, 486}

\bibitem[\protect\citeauthoryear{{Conroy}, {White}  \& {Gunn}}{{Conroy}
  et~al.}{2010}]{Conroy+10}
{Conroy} C.,  {White} M.,   {Gunn} J.~E.,  2010, \mn@doi [\apj]
  {10.1088/0004-637X/708/1/58}, \href
  {https://ui.adsabs.harvard.edu/abs/2010ApJ...708...58C} {708, 58}

\bibitem[\protect\citeauthoryear{{Cowie}, {Songaila}, {Hu}  \& {Cohen}}{{Cowie}
  et~al.}{1996}]{Cowie+96}
{Cowie} L.~L.,  {Songaila} A.,  {Hu} E.~M.,   {Cohen} J.~G.,  1996, \mn@doi
  [\aj] {10.1086/118058}, \href
  {https://ui.adsabs.harvard.edu/abs/1996AJ....112..839C} {112, 839}

\bibitem[\protect\citeauthoryear{{Das}, {Pandey}  \& {Sarkar}}{{Das}
  et~al.}{2021}]{Das+21}
{Das} A.,  {Pandey} B.,   {Sarkar} S.,  2021, arXiv e-prints, \href
  {https://ui.adsabs.harvard.edu/abs/2021arXiv210102564D} {p. arXiv:2101.02564}

\bibitem[\protect\citeauthoryear{{Dom{\'\i}nguez S{\'a}nchez},
  {Huertas-Company}, {Bernardi}, {Tuccillo}  \& {Fischer}}{{Dom{\'\i}nguez
  S{\'a}nchez} et~al.}{2018}]{DominguezSanchez+18}
{Dom{\'\i}nguez S{\'a}nchez} H.,  {Huertas-Company} M.,  {Bernardi} M.,
  {Tuccillo} D.,   {Fischer} J.~L.,  2018, \mn@doi [\mnras]
  {10.1093/mnras/sty338}, \href
  {https://ui.adsabs.harvard.edu/abs/2018MNRAS.476.3661D} {476, 3661}

\bibitem[\protect\citeauthoryear{{Drory} et~al.,}{{Drory}
  et~al.}{2015}]{Drory+15}
{Drory} N.,  et~al., 2015, \mn@doi [AJ] {10.1088/0004-6256/149/2/77}, \href
  {http://adsabs.harvard.edu/abs/2015AJ....149...77D} {149, 77}

\bibitem[\protect\citeauthoryear{{Emsellem} et~al.,}{{Emsellem}
  et~al.}{2007}]{Emsellem+07}
{Emsellem} E.,  et~al., 2007, \mn@doi [\mnras]
  {10.1111/j.1365-2966.2007.11752.x}, \href
  {https://ui.adsabs.harvard.edu/abs/2007MNRAS.379..401E} {379, 401}

\bibitem[\protect\citeauthoryear{{Fang}, {Faber}, {Koo}  \& {Dekel}}{{Fang}
  et~al.}{2013}]{Fang+13}
{Fang} J.~J.,  {Faber} S.~M.,  {Koo} D.~C.,   {Dekel} A.,  2013, \mn@doi [\apj]
  {10.1088/0004-637X/776/1/63}, \href
  {https://ui.adsabs.harvard.edu/abs/2013ApJ...776...63F} {776, 63}

\bibitem[\protect\citeauthoryear{{Fischer}, {Dom{\'\i}nguez S{\'a}nchez}  \&
  {Bernardi}}{{Fischer} et~al.}{2019}]{Fischer+19}
{Fischer} J.~L.,  {Dom{\'\i}nguez S{\'a}nchez} H.,   {Bernardi} M.,  2019,
  \mn@doi [\mnras] {10.1093/mnras/sty3135}, \href
  {https://ui.adsabs.harvard.edu/abs/2019MNRAS.483.2057F} {483, 2057}

\bibitem[\protect\citeauthoryear{{Fontanot}, {De Lucia}, {Monaco}, {Somerville}
   \& {Santini}}{{Fontanot} et~al.}{2009}]{Fontanot+09}
{Fontanot} F.,  {De Lucia} G.,  {Monaco} P.,  {Somerville} R.~S.,   {Santini}
  P.,  2009, \mn@doi [\mnras] {10.1111/j.1365-2966.2009.15058.x}, \href
  {https://ui.adsabs.harvard.edu/abs/2009MNRAS.397.1776F} {397, 1776}

\bibitem[\protect\citeauthoryear{{Fraser-McKelvie} et~al.,}{{Fraser-McKelvie}
  et~al.}{2019}]{FraserMcKelvie+19}
{Fraser-McKelvie} A.,  et~al., 2019, \mn@doi [\mnras] {10.1093/mnrasl/slz085},
  \href {https://ui.adsabs.harvard.edu/abs/2019MNRAS.488L...6F} {488, L6}

\bibitem[\protect\citeauthoryear{{Freeman}}{{Freeman}}{1970}]{Freeman70}
{Freeman} K.~C.,  1970, \mn@doi [\apj] {10.1086/150474}, \href
  {https://ui.adsabs.harvard.edu/abs/1970ApJ...160..811F} {160, 811}

\bibitem[\protect\citeauthoryear{{Garc{\'\i}a-Benito}
  et~al.,}{{Garc{\'\i}a-Benito} et~al.}{2017}]{GarciaBenito+17}
{Garc{\'\i}a-Benito} R.,  et~al., 2017, \mn@doi [\aap]
  {10.1051/0004-6361/201731357}, \href
  {https://ui.adsabs.harvard.edu/abs/2017A&A...608A..27G} {608, A27}

\bibitem[\protect\citeauthoryear{{Garc{\'\i}a-Benito}, {Gonz{\'a}lez Delgado},
  {P{\'e}rez}, {Cid Fernandes}, {S{\'a}nchez}  \& {de
  Amorim}}{{Garc{\'\i}a-Benito} et~al.}{2019}]{GarciaBenito+19}
{Garc{\'\i}a-Benito} R.,  {Gonz{\'a}lez Delgado} R.~M.,  {P{\'e}rez} E.,  {Cid
  Fernandes} R.,  {S{\'a}nchez} S.~F.,   {de Amorim} A.~L.,  2019, \mn@doi
  [\aap] {10.1051/0004-6361/201833993}, \href
  {https://ui.adsabs.harvard.edu/abs/2019A&A...621A.120G} {621, A120}

\bibitem[\protect\citeauthoryear{{Ge}, {Yan}, {Cappellari}, {Mao}, {Li}  \&
  {Lu}}{{Ge} et~al.}{2018}]{Ge+18}
{Ge} J.,  {Yan} R.,  {Cappellari} M.,  {Mao} S.,  {Li} H.,   {Lu} Y.,  2018,
  \mn@doi [\mnras] {10.1093/mnras/sty1245}, \href
  {http://adsabs.harvard.edu/abs/2018MNRAS.478.2633G} {478, 2633}

\bibitem[\protect\citeauthoryear{{Ge}, {Mao}, {Lu}, {Cappellari}  \&
  {Yan}}{{Ge} et~al.}{2019}]{Ge+19}
{Ge} J.,  {Mao} S.,  {Lu} Y.,  {Cappellari} M.,   {Yan} R.,  2019, \mn@doi
  [\mnras] {10.1093/mnras/stz418}, \href
  {https://ui.adsabs.harvard.edu/abs/2019MNRAS.485.1675G} {485, 1675}

\bibitem[\protect\citeauthoryear{{Goddard} et~al.,}{{Goddard}
  et~al.}{2017}]{Goddard+17}
{Goddard} D.,  et~al., 2017, \mn@doi [\mnras] {10.1093/mnras/stw3371}, \href
  {http://adsabs.harvard.edu/abs/2017MNRAS.466.4731G} {466, 4731}

\bibitem[\protect\citeauthoryear{{Gonz{\'a}lez Delgado} et~al.,}{{Gonz{\'a}lez
  Delgado} et~al.}{2016}]{GonzalezDelgado+16}
{Gonz{\'a}lez Delgado} R.~M.,  et~al., 2016, \mn@doi [\aap]
  {10.1051/0004-6361/201628174}, \href
  {https://ui.adsabs.harvard.edu/abs/2016A&A...590A..44G} {590, A44}

\bibitem[\protect\citeauthoryear{{Gonz{\'a}lez Delgado} et~al.,}{{Gonz{\'a}lez
  Delgado} et~al.}{2017}]{GonzalezDelgado+17}
{Gonz{\'a}lez Delgado} R.~M.,  et~al., 2017, \mn@doi [\aap]
  {10.1051/0004-6361/201730883}, \href
  {http://adsabs.harvard.edu/abs/2017A%26A...607A.128G} {607, A128}

\bibitem[\protect\citeauthoryear{{Graham} et~al.,}{{Graham}
  et~al.}{2018}]{Graham+18}
{Graham} M.~T.,  et~al., 2018, \mn@doi [\mnras] {10.1093/mnras/sty504}, \href
  {https://ui.adsabs.harvard.edu/abs/2018MNRAS.477.4711G} {477, 4711}

\bibitem[\protect\citeauthoryear{{Greener} et~al.,}{{Greener}
  et~al.}{2020}]{Greener+20}
{Greener} M.~J.,  et~al., 2020, \mn@doi [\mnras] {10.1093/mnras/staa1300},
  \href {https://ui.adsabs.harvard.edu/abs/2020MNRAS.tmp.1487G} {}

\bibitem[\protect\citeauthoryear{{Gunn} et~al.,}{{Gunn} et~al.}{2006}]{Gunn+06}
{Gunn} J.~E.,  et~al., 2006, \mn@doi [AJ] {10.1086/500975}, \href
  {http://adsabs.harvard.edu/abs/2006AJ....131.2332G} {131, 2332}

\bibitem[\protect\citeauthoryear{{Hart} et~al.,}{{Hart} et~al.}{2016}]{Hart+16}
{Hart} R.~E.,  et~al., 2016, \mn@doi [MNRAS] {10.1093/mnras/stw1588}, \href
  {http://adsabs.harvard.edu/abs/2016MNRAS.461.3663H} {461, 3663}

\bibitem[\protect\citeauthoryear{{Hart} et~al.,}{{Hart} et~al.}{2017}]{Hart+17}
{Hart} R.~E.,  et~al., 2017, \mn@doi [MNRAS] {10.1093/mnras/stx2137}, \href
  {http://adsabs.harvard.edu/abs/2017MNRAS.472.2263H} {472, 2263}

\bibitem[\protect\citeauthoryear{{Hart}, {Bamford}, {Keel}, {Kruk}, {Masters},
  {Simmons}  \& {Smethurst}}{{Hart} et~al.}{2018}]{Hart+18}
{Hart} R.~E.,  {Bamford} S.~P.,  {Keel} W.~C.,  {Kruk} S.~J.,  {Masters} K.~L.,
   {Simmons} B.~D.,   {Smethurst} R.~J.,  2018, \mn@doi [\mnras]
  {10.1093/mnras/sty1201}, \href
  {https://ui.adsabs.harvard.edu/abs/2018MNRAS.478..932H} {478, 932}

\bibitem[\protect\citeauthoryear{{Heavens}, {Panter}, {Jimenez}  \&
  {Dunlop}}{{Heavens} et~al.}{2004}]{Heavens+04}
{Heavens} A.,  {Panter} B.,  {Jimenez} R.,   {Dunlop} J.,  2004, \mn@doi [\nat]
  {10.1038/nature02474}, \href
  {https://ui.adsabs.harvard.edu/abs/2004Natur.428..625H} {428, 625}

\bibitem[\protect\citeauthoryear{{Holmberg}}{{Holmberg}}{1958}]{Holmberg58}
{Holmberg} E.,  1958, Meddelanden fran Lunds Astronomiska Observatorium Serie
  II, \href {https://ui.adsabs.harvard.edu/abs/1958MeLuS.136....1H} {136, 1}

\bibitem[\protect\citeauthoryear{{Hopkins} \& {Beacom}}{{Hopkins} \&
  {Beacom}}{2006}]{HopkinsBeacom06}
{Hopkins} A.~M.,  {Beacom} J.~F.,  2006, \mn@doi [\apj] {10.1086/506610}, \href
  {https://ui.adsabs.harvard.edu/abs/2006ApJ...651..142H} {651, 142}

\bibitem[\protect\citeauthoryear{{Ibarra-Medel} et~al.,}{{Ibarra-Medel}
  et~al.}{2016}]{IbarraMedel+16}
{Ibarra-Medel} H.~J.,  et~al., 2016, \mn@doi [\mnras] {10.1093/mnras/stw2126},
  \href {http://adsabs.harvard.edu/abs/2016MNRAS.463.2799I} {463, 2799}

\bibitem[\protect\citeauthoryear{{Kennicutt}}{{Kennicutt}}{1983}]{Kennicutt83}
{Kennicutt} R.~C. J.,  1983, \mn@doi [\apj] {10.1086/161261}, \href
  {https://ui.adsabs.harvard.edu/abs/1983ApJ...272...54K} {272, 54}

\bibitem[\protect\citeauthoryear{{Kennicutt}}{{Kennicutt}}{1998}]{Kennicutt98}
{Kennicutt} Jr. R.~C.,  1998, \mn@doi [ARA\&A]
  {10.1146/annurev.astro.36.1.189}, \href
  {http://adsabs.harvard.edu/abs/1998ARA%26A..36..189K} {36, 189}

\bibitem[\protect\citeauthoryear{{Lacerna}, {Ibarra-Medel}, {Avila-Reese},
  {Hern{\'a}ndez-Toledo}, {V{\'a}zquez-Mata}  \& {S{\'a}nchez}}{{Lacerna}
  et~al.}{2020}]{Lacerna+20}
{Lacerna} I.,  {Ibarra-Medel} H.,  {Avila-Reese} V.,  {Hern{\'a}ndez-Toledo}
  H.~M.,  {V{\'a}zquez-Mata} J.~A.,   {S{\'a}nchez} S.~F.,  2020, \mn@doi
  [\aap] {10.1051/0004-6361/202037503}, \href
  {https://ui.adsabs.harvard.edu/abs/2020A&A...644A.117L} {644, A117}

\bibitem[\protect\citeauthoryear{{Law} et~al.,}{{Law} et~al.}{2015}]{Law+15}
{Law} D.~R.,  et~al., 2015, \mn@doi [AJ] {10.1088/0004-6256/150/1/19}, \href
  {http://adsabs.harvard.edu/abs/2015AJ....150...19L} {150, 19}

\bibitem[\protect\citeauthoryear{{Law} et~al.,}{{Law} et~al.}{2016}]{Law+16}
{Law} D.~R.,  et~al., 2016, \mn@doi [AJ] {10.3847/0004-6256/152/4/83}, \href
  {http://adsabs.harvard.edu/abs/2016AJ....152...83L} {152, 83}

\bibitem[\protect\citeauthoryear{{Lee}, {Worthey}, {Trager}  \& {Faber}}{{Lee}
  et~al.}{2007}]{Lee+07}
{Lee} H.-c.,  {Worthey} G.,  {Trager} S.~C.,   {Faber} S.~M.,  2007, \mn@doi
  [\apj] {10.1086/518855}, \href
  {https://ui.adsabs.harvard.edu/abs/2007ApJ...664..215L} {664, 215}

\bibitem[\protect\citeauthoryear{{Li} et~al.,}{{Li} et~al.}{2017}]{Li+17}
{Li} H.,  et~al., 2017, \mn@doi [\apj] {10.3847/1538-4357/aa662a}, \href
  {https://ui.adsabs.harvard.edu/abs/2017ApJ...838...77L} {838, 77}

\bibitem[\protect\citeauthoryear{{Lintott} et~al.,}{{Lintott}
  et~al.}{2008}]{Lintott+08}
{Lintott} C.~J.,  et~al., 2008, \mn@doi [MNRAS]
  {10.1111/j.1365-2966.2008.13689.x}, \href
  {http://adsabs.harvard.edu/abs/2008MNRAS.389.1179L} {389, 1179}

\bibitem[\protect\citeauthoryear{{L{\'o}pez Fern{\'a}ndez} et~al.,}{{L{\'o}pez
  Fern{\'a}ndez} et~al.}{2018}]{LopezFernandez+18}
{L{\'o}pez Fern{\'a}ndez} R.,  et~al., 2018, \mn@doi [\aap]
  {10.1051/0004-6361/201732358}, \href
  {https://ui.adsabs.harvard.edu/abs/2018A&A...615A..27L} {615, A27}

\bibitem[\protect\citeauthoryear{{Madau} \& {Dickinson}}{{Madau} \&
  {Dickinson}}{2014}]{MD14}
{Madau} P.,  {Dickinson} M.,  2014, \mn@doi [\araa]
  {10.1146/annurev-astro-081811-125615}, \href
  {https://ui.adsabs.harvard.edu/abs/2014ARA&A..52..415M} {52, 415}

\bibitem[\protect\citeauthoryear{{Madau}, {Ferguson}, {Dickinson},
  {Giavalisco}, {Steidel}  \& {Fruchter}}{{Madau} et~al.}{1996}]{Madau+96}
{Madau} P.,  {Ferguson} H.~C.,  {Dickinson} M.~E.,  {Giavalisco} M.,  {Steidel}
  C.~C.,   {Fruchter} A.,  1996, \mn@doi [\mnras] {10.1093/mnras/283.4.1388},
  \href {https://ui.adsabs.harvard.edu/abs/1996MNRAS.283.1388M} {283, 1388}

\bibitem[\protect\citeauthoryear{{Maraston}}{{Maraston}}{1998}]{Maraston98}
{Maraston} C.,  1998, \mn@doi [\mnras] {10.1046/j.1365-8711.1998.01947.x},
  \href {http://adsabs.harvard.edu/abs/1998MNRAS.300..872M} {300, 872}

\bibitem[\protect\citeauthoryear{{Maraston}}{{Maraston}}{2005}]{Maraston05}
{Maraston} C.,  2005, \mn@doi [\mnras] {10.1111/j.1365-2966.2005.09270.x},
  \href {https://ui.adsabs.harvard.edu/abs/2005MNRAS.362..799M} {362, 799}

\bibitem[\protect\citeauthoryear{{Martig}, {Bournaud}, {Teyssier}  \&
  {Dekel}}{{Martig} et~al.}{2009}]{Martig+09}
{Martig} M.,  {Bournaud} F.,  {Teyssier} R.,   {Dekel} A.,  2009, \mn@doi
  [\apj] {10.1088/0004-637X/707/1/250}, \href
  {https://ui.adsabs.harvard.edu/abs/2009ApJ...707..250M} {707, 250}

\bibitem[\protect\citeauthoryear{{Masters} et~al.,}{{Masters}
  et~al.}{2019}]{Masters+19}
{Masters} K.~L.,  et~al., 2019, \mn@doi [\mnras] {10.1093/mnras/stz1153}, \href
  {https://ui.adsabs.harvard.edu/abs/2019MNRAS.487.1808M} {487, 1808}

\bibitem[\protect\citeauthoryear{{Moresco} et~al.,}{{Moresco}
  et~al.}{2013}]{Moresco+13}
{Moresco} M.,  et~al., 2013, \mn@doi [\aap] {10.1051/0004-6361/201321797},
  \href {https://ui.adsabs.harvard.edu/abs/2013A&A...558A..61M} {558, A61}

\bibitem[\protect\citeauthoryear{{Mortlock} et~al.,}{{Mortlock}
  et~al.}{2013}]{Mortlock+13}
{Mortlock} A.,  et~al., 2013, \mn@doi [\mnras] {10.1093/mnras/stt793}, \href
  {https://ui.adsabs.harvard.edu/abs/2013MNRAS.433.1185M} {433, 1185}

\bibitem[\protect\citeauthoryear{{Muzzin} et~al.,}{{Muzzin}
  et~al.}{2013}]{Muzzin+13}
{Muzzin} A.,  et~al., 2013, \mn@doi [\apj] {10.1088/0004-637X/777/1/18}, \href
  {https://ui.adsabs.harvard.edu/abs/2013ApJ...777...18M} {777, 18}

\bibitem[\protect\citeauthoryear{{Naim} et~al.,}{{Naim} et~al.}{1995}]{Naim+95}
{Naim} A.,  et~al., 1995, \mn@doi [\mnras] {10.1093/mnras/274.4.1107}, \href
  {https://ui.adsabs.harvard.edu/abs/1995MNRAS.274.1107N} {274, 1107}

\bibitem[\protect\citeauthoryear{{Nair} \& {Abraham}}{{Nair} \&
  {Abraham}}{2010}]{NairAbraham10}
{Nair} P.~B.,  {Abraham} R.~G.,  2010, \mn@doi [\apjs]
  {10.1088/0067-0049/186/2/427}, \href
  {https://ui.adsabs.harvard.edu/abs/2010ApJS..186..427N} {186, 427}

\bibitem[\protect\citeauthoryear{{Osterbrock} \& {Ferland}}{{Osterbrock} \&
  {Ferland}}{2006}]{OsterbrockFerland06}
{Osterbrock} D.~E.,  {Ferland} G.~J.,  2006, {Astrophysics of gaseous nebulae
  and active galactic nuclei}.
{University Science Books}

\bibitem[\protect\citeauthoryear{{Panter}, {Heavens}  \& {Jimenez}}{{Panter}
  et~al.}{2003}]{PanterHeavensJimenez03}
{Panter} B.,  {Heavens} A.~F.,   {Jimenez} R.,  2003, \mn@doi [\mnras]
  {10.1046/j.1365-8711.2003.06722.x}, \href
  {http://adsabs.harvard.edu/abs/2003MNRAS.343.1145P} {343, 1145}

\bibitem[\protect\citeauthoryear{{Panter}, {Jimenez}, {Heavens}  \&
  {Charlot}}{{Panter} et~al.}{2007}]{Panter+07}
{Panter} B.,  {Jimenez} R.,  {Heavens} A.~F.,   {Charlot} S.,  2007, \mn@doi
  [\mnras] {10.1111/j.1365-2966.2007.11909.x}, \href
  {http://adsabs.harvard.edu/abs/2007MNRAS.378.1550P} {378, 1550}

\bibitem[\protect\citeauthoryear{{Peng} et~al.,}{{Peng} et~al.}{2010}]{Peng+10}
{Peng} Y.-j.,  et~al., 2010, \mn@doi [\apj] {10.1088/0004-637X/721/1/193},
  \href {https://ui.adsabs.harvard.edu/abs/2010ApJ...721..193P} {721, 193}

\bibitem[\protect\citeauthoryear{{P{\'e}rez} et~al.,}{{P{\'e}rez}
  et~al.}{2013}]{Perez+13}
{P{\'e}rez} E.,  et~al., 2013, \mn@doi [\apjl] {10.1088/2041-8205/764/1/L1},
  \href {https://ui.adsabs.harvard.edu/abs/2013ApJ...764L...1P} {764, L1}

\bibitem[\protect\citeauthoryear{{Peterken} et~al.,}{{Peterken}
  et~al.}{2019}]{Peterken+19TS}
{Peterken} T.~G.,  et~al., 2019, \mn@doi [\mnras] {10.1093/mnras/stz2204},
  \href {https://ui.adsabs.harvard.edu/doi/10.1093/mnras/stz2204} {489, 1338}

\bibitem[\protect\citeauthoryear{{Peterken}, {Merrifield},
  {Arag{\'o}n-Salamanca}, {Fraser-McKelvie}, {Avila-Reese}, {Riffel}, {Knapen}
  \& {Drory}}{{Peterken} et~al.}{2020}]{Peterken+20FR}
{Peterken} T.,  {Merrifield} M.,  {Arag{\'o}n-Salamanca} A.,  {Fraser-McKelvie}
  A.,  {Avila-Reese} V.,  {Riffel} R.,  {Knapen} J.,   {Drory} N.,  2020,
  \mn@doi [\mnras] {10.1093/mnras/staa1303}, \href
  {https://ui.adsabs.harvard.edu/abs/2020MNRAS.495.3387P} {495, 3387}

\bibitem[\protect\citeauthoryear{{Peterken}, {Merrifield},
  {Arag{\'o}n-Salamanca}, {Avila-Reese}, {Boardman}, {Drory}  \&
  {Lane}}{{Peterken} et~al.}{2021}]{Peterken+20Morph}
{Peterken} T.,  {Merrifield} M.,  {Arag{\'o}n-Salamanca} A.,  {Avila-Reese} V.,
   {Boardman} N.~F.,  {Drory} N.,   {Lane} R.~R.,  2021, \mn@doi [\mnras]
  {10.1093/mnrasl/slaa179}, \href
  {https://ui.adsabs.harvard.edu/abs/2021MNRAS.500L..42P} {500, L42}

\bibitem[\protect\citeauthoryear{{Pforr}, {Maraston}  \& {Tonini}}{{Pforr}
  et~al.}{2012}]{Pforr+12}
{Pforr} J.,  {Maraston} C.,   {Tonini} C.,  2012, \mn@doi [\mnras]
  {10.1111/j.1365-2966.2012.20848.x}, \href
  {https://ui.adsabs.harvard.edu/abs/2012MNRAS.422.3285P} {422, 3285}

\bibitem[\protect\citeauthoryear{{Planck Collaboration} et~al.,}{{Planck
  Collaboration} et~al.}{2016}]{Planck15}
{Planck Collaboration} et~al., 2016, \mn@doi [\aap]
  {10.1051/0004-6361/201525830}, \href
  {https://ui.adsabs.harvard.edu/abs/2016A&A...594A..13P} {594, A13}

\bibitem[\protect\citeauthoryear{{Roberts}}{{Roberts}}{1963}]{Roberts63}
{Roberts} M.~S.,  1963, \mn@doi [\araa] {10.1146/annurev.aa.01.090163.001053},
  \href {https://ui.adsabs.harvard.edu/abs/1963ARA&A...1..149R} {1, 149}

\bibitem[\protect\citeauthoryear{{Rodriguez-Puebla}, {Calette}, {Avila-Reese},
  {Rodriguez-Gomez}  \& {Huertas-Company}}{{Rodriguez-Puebla}
  et~al.}{2020}]{RodriguezPuebla+20}
{Rodriguez-Puebla} A.,  {Calette} A.~R.,  {Avila-Reese} V.,  {Rodriguez-Gomez}
  V.,   {Huertas-Company} M.,  2020, arXiv e-prints, \href
  {https://ui.adsabs.harvard.edu/abs/2020arXiv200413740R} {p. arXiv:2004.13740}

\bibitem[\protect\citeauthoryear{{S{\'a}nchez} et~al.,}{{S{\'a}nchez}
  et~al.}{2016}]{Pipe3D}
{S{\'a}nchez} S.~F.,  et~al., 2016, \rmxaa, \href
  {http://adsabs.harvard.edu/abs/2016RMxAA..52...21S} {52, 21}

\bibitem[\protect\citeauthoryear{{S{\'a}nchez} et~al.,}{{S{\'a}nchez}
  et~al.}{2019}]{Sanchez+19}
{S{\'a}nchez} S.~F.,  et~al., 2019, \mn@doi [\mnras] {10.1093/mnras/sty2730},
  \href {http://adsabs.harvard.edu/abs/2019MNRAS.482.1557S} {482, 1557}

\bibitem[\protect\citeauthoryear{{Schawinski} et~al.,}{{Schawinski}
  et~al.}{2014}]{Schawinski+14}
{Schawinski} K.,  et~al., 2014, \mn@doi [\mnras] {10.1093/mnras/stu327}, \href
  {https://ui.adsabs.harvard.edu/abs/2014MNRAS.440..889S} {440, 889}

\bibitem[\protect\citeauthoryear{{Smee} et~al.,}{{Smee} et~al.}{2013}]{Smee+13}
{Smee} S.~A.,  et~al., 2013, \mn@doi [AJ] {10.1088/0004-6256/146/2/32}, \href
  {http://adsabs.harvard.edu/abs/2013AJ....146...32S} {146, 32}

\bibitem[\protect\citeauthoryear{{Smethurst} et~al.,}{{Smethurst}
  et~al.}{2015}]{Smethurst+15}
{Smethurst} R.~J.,  et~al., 2015, \mn@doi [\mnras] {10.1093/mnras/stv161},
  \href {https://ui.adsabs.harvard.edu/abs/2015MNRAS.450..435S} {450, 435}

\bibitem[\protect\citeauthoryear{{Smethurst}, {Lintott}, {Bamford}, {Hart},
  {Kruk}, {Masters}, {Nichol}  \& {Simmons}}{{Smethurst}
  et~al.}{2017}]{Smethurst+17}
{Smethurst} R.~J.,  {Lintott} C.~J.,  {Bamford} S.~P.,  {Hart} R.~E.,  {Kruk}
  S.~J.,  {Masters} K.~L.,  {Nichol} R.~C.,   {Simmons} B.~D.,  2017, \mn@doi
  [\mnras] {10.1093/mnras/stx973}, \href
  {https://ui.adsabs.harvard.edu/abs/2017MNRAS.469.3670S} {469, 3670}

\bibitem[\protect\citeauthoryear{{Smethurst} et~al.,}{{Smethurst}
  et~al.}{2018}]{Smethurst+18}
{Smethurst} R.~J.,  et~al., 2018, \mn@doi [\mnras] {10.1093/mnras/stx2547},
  \href {https://ui.adsabs.harvard.edu/abs/2018MNRAS.473.2679S} {473, 2679}

\bibitem[\protect\citeauthoryear{{Vazdekis}, {Koleva}, {Ricciardelli},
  {R{\"o}ck}  \& {Falc{\'o}n-Barroso}}{{Vazdekis} et~al.}{2016}]{E-MILES}
{Vazdekis} A.,  {Koleva} M.,  {Ricciardelli} E.,  {R{\"o}ck} B.,
  {Falc{\'o}n-Barroso} J.,  2016, \mn@doi [\mnras] {10.1093/mnras/stw2231},
  \href {https://ui.adsabs.harvard.edu/\#abs/2016MNRAS.463.3409V} {463, 3409}

\bibitem[\protect\citeauthoryear{{Wake} et~al.,}{{Wake} et~al.}{2017}]{Wake+17}
{Wake} D.~A.,  et~al., 2017, \mn@doi [AJ] {10.3847/1538-3881/aa7ecc}, \href
  {http://adsabs.harvard.edu/abs/2017AJ....154...86W} {154, 86}

\bibitem[\protect\citeauthoryear{{Wang}, {Cappellari}, {Peng}  \&
  {Graham}}{{Wang} et~al.}{2020}]{Wang+20}
{Wang} B.,  {Cappellari} M.,  {Peng} Y.,   {Graham} M.,  2020, \mn@doi [\mnras]
  {10.1093/mnras/staa1325}, \href
  {https://ui.adsabs.harvard.edu/abs/2020MNRAS.495.1958W} {495, 1958}

\bibitem[\protect\citeauthoryear{{Westfall} et~al.,}{{Westfall}
  et~al.}{2019}]{DAP}
{Westfall} K.~B.,  et~al., 2019, \mn@doi [\aj] {10.3847/1538-3881/ab44a2},
  \href {https://ui.adsabs.harvard.edu/abs/2019AJ....158..231W} {158, 231}

\bibitem[\protect\citeauthoryear{{Willett} et~al.,}{{Willett}
  et~al.}{2013}]{Willett+13}
{Willett} K.~W.,  et~al., 2013, \mn@doi [MNRAS] {10.1093/mnras/stt1458}, \href
  {http://adsabs.harvard.edu/abs/2013MNRAS.435.2835W} {435, 2835}

\bibitem[\protect\citeauthoryear{{Wuyts} et~al.,}{{Wuyts}
  et~al.}{2011}]{Wuyts+11}
{Wuyts} S.,  et~al., 2011, \mn@doi [\apj] {10.1088/0004-637X/742/2/96}, \href
  {https://ui.adsabs.harvard.edu/abs/2011ApJ...742...96W} {742, 96}

\bibitem[\protect\citeauthoryear{{Yan} et~al.,}{{Yan}
  et~al.}{2016a}]{Yan+16-cal}
{Yan} R.,  et~al., 2016a, \mn@doi [AJ] {10.3847/0004-6256/151/1/8}, \href
  {http://adsabs.harvard.edu/abs/2016AJ....151....8Y} {151, 8}

\bibitem[\protect\citeauthoryear{{Yan} et~al.,}{{Yan}
  et~al.}{2016b}]{Yan+16-design}
{Yan} R.,  et~al., 2016b, \mn@doi [AJ] {10.3847/0004-6256/152/6/197}, \href
  {http://adsabs.harvard.edu/abs/2016AJ....152..197Y} {152, 197}

\bibitem[\protect\citeauthoryear{{Yi}}{{Yi}}{2003}]{Yi03}
{Yi} S.~K.,  2003, \mn@doi [\apj] {10.1086/344640}, \href
  {https://ui.adsabs.harvard.edu/abs/2003ApJ...582..202Y} {582, 202}

\bibitem[\protect\citeauthoryear{{Yi}}{{Yi}}{2008}]{Yi08}
{Yi} S.~K.,  2008, in {Heber} U.,  {Jeffery} C.~S.,   {Napiwotzki} R.,  eds,
  Astronomical Society of the Pacific Conference Series Vol. 392, Hot Subdwarf
  Stars and Related Objects. Astronomical Society of the Pacific, p.~3

\bibitem[\protect\citeauthoryear{de {Amorim} et~al.,}{de~{Amorim}
  et~al.}{2017}]{deAmorim+17}
de {Amorim} A.~L.,  et~al., 2017, \mn@doi [\mnras] {10.1093/mnras/stx1805},
  \href {https://ui.adsabs.harvard.edu/abs/2017MNRAS.471.3727D} {471, 3727}

\bibitem[\protect\citeauthoryear{{van Dokkum} et~al.,}{{van Dokkum}
  et~al.}{2008}]{vanDokkum+08}
{van Dokkum} P.~G.,  et~al., 2008, \mn@doi [\apjl] {10.1086/587874}, \href
  {https://ui.adsabs.harvard.edu/abs/2008ApJ...677L...5V} {677, L5}

\makeatother
\end{thebibliography}




\bsp	
\label{lastpage}
\end{document}